\documentclass[a4paper,fleqn,usenatbib]{mnras}
\usepackage[T1]{fontenc}
\usepackage{ae,aecompl}

\usepackage{graphicx}   
\usepackage{amsmath}    
\usepackage{amssymb}    

\usepackage{bm}
\usepackage{multirow}
\usepackage{ifthen}

\usepackage{lineno}

\newcommand{\AU}{\mathrm{AU}}
\newcommand{\Msun}{M_{\odot}}
\newcommand{\yr}{\mathrm{yr}}
\newcommand{\Kelvin}{\mathrm{K}}
\newcommand{\kilo}{\mathrm{k}}
\newcommand{\mega}{\mathrm{M}}

\newcommand{\primary}{\mathrm{p}}
\newcommand{\secondary}{\mathrm{s}}
\newcommand{\thebinary}{\mathrm{b}}
\newcommand{\system}{\mathrm{sys}}
\newcommand{\disc}{\mathrm{d}}
\newcommand{\tidal}{\mathrm{t}}
\newcommand{\encounter}{\mathrm{e}}

\newcommand{\dee}[1]{\mathrm{d}{#1}}
\newcommand{\rbrackets}[1]{\left(#1\right)}     
\newcommand{\sbrackets}[1]{\left[#1\right]}     
\newcommand{\abrackets}[1]{\left<#1\right>}     
\newcommand{\vbrackets}[1]{\left|#1\right|}     

\title[Discs in misaligned binary systems]{Discs in misaligned binary systems}

\author[K. Rawiraswattana, D. A. Hubber and S. P. Goodwin]{
    Krisada Rawiraswattana$^{1}$\thanks{E-mail: krisada.r@psu.ac.th},
    David A. Hubber$^{2,3}$ and
    Simon P. Goodwin$^{4}$\\
    $^{1}$Department of Physics, Faculty of Science, Prince of Songkla University, Hatyai, Songkla 90110, Thailand\\
    $^{2}$University Observatory, Ludwig-Maximilians-University Munich, Scheinerstr. 1, D-81679 Munich, Germany\\
    $^{3}$Excellence Cluster Universe, Boltzmannstr. 2, D-85748 Garching, Germany\\
    $^{4}$Department of Physics and Astronomy, University of Sheffield, Sheffield S3 7RH, UK
}

\date{Accepted XXX. Received YYY; in original form ZZZ}
\pubyear{2015}

\begin{document}
\label{firstpage}
\pagerange{\pageref{firstpage}--\pageref{lastpage}}
\maketitle

\begin{abstract}
We perform SPH simulations to study precession and changes in alignment between the circumprimary disc and the binary orbit in misaligned binary systems.
We find that the precession process can be described by the rigid-disc approximation, where the disc is considered as a rigid body interacting with the binary companion only gravitationally.
Precession also causes change in alignment between the rotational axis of the disc and the spin axis of the primary star.
This type of alignment is of great important for explaining the origin of spin-orbit misaligned planetary systems.
However, we find that the rigid-disc approximation fails to describe changes in alignment between the disc and the binary orbit.
This is because the alignment process is a consequence of interactions that involve the fluidity of the disc, such as the tidal interaction and the encounter interaction.
Furthermore, simulation results show that there are not only alignment processes, which bring the components towards alignment, but also anti-alignment processes, which tend to misalign the components.
The alignment process dominates in systems with misalignment angle near $90\degr$, while the anti-alignment process dominates in systems with the misalignment angle near $0\degr$ or $180\degr$.
This means that highly misaligned systems will become more aligned but slightly misaligned systems will become more misaligned.
\end{abstract}


\section{Introduction}

Young stars are often observed to have discs of gas and dust which are the remnants of the star formation process and the sites of planet formation.

Many planetary systems are observed to be `misaligned', that is that the rotational axis of the star and the orbital axis of the planetary system are different \citep{Winn:etal:2009a,Winn:etal:2009b,Winn:etal:2010,Batygin:2012}.
This may be due to disc forming misaligned \citep[e.g.][]{Tremaine:1991,Walch:etal:2010,Bate:etal:2010,Fielding:etal:2015}, or due to close encounters \citep[e.g.][]{Thies:etal:2011,Rosotti:etal:2014}, or due to migration \citep[e.g.][]{Fabrycky:Tremaine:2007,Nagasawa:etal:2008}, or may be due to magnetic torques \citep{Lai:etal:2011}.

Many (perhaps the vast majority) of young stars are in multiple systems \citep[e.g.][]{Mathieu:1994,Patience:etal:2002,King:etal:2012,Duchene:Kraus:2013}.
A number of these young binaries have been found with misaligned discs \citep[e.g.][and references therein]{Monin:etal:2006}.
In the Taurus-Auriga and Scorpius-Ophiuchus star-forming regions, mildly misaligned discs with misalignment angles $\lesssim 20\degr$ are found in wide T Tauri binaries with separations between $200-1000\AU$ \citep{Jensen:etal:2004}.
As an example from some resolved systems, the protobinary system HH 24 MMS with separation $\sim 360\AU$ has misaligned discs around the components with a relatively large difference in position angles \citep[$\sim 45\degr$,][]{Kang:etal:2008}.
The discs surrounding the components in Haro 6-10, a T Tauri binary system with separation $\sim 160\AU$, are seen to be strongly misaligned with each other by $\sim 70\degr$ \citep{Roccatagliata:etal:2011}.
Indeed, the discs surrounding the archetypal T Tauri triple (separation between T Tau N and T Tau Sab $\gtrsim 100\AU$) are also found to be relatively misaligned to each other \citep{Skemer:etal:2008,Ratzka:etal:2009}.
From these observations, the components in misaligned systems seem to have mutual separations $\gtrsim 100\AU$.

Misaligned disc-binary systems may form primordially in turbulent environments \citep[in the same way as misaligned discs above, e.g.][]{Bate:etal:2003,Goodwin:etal:2004a,Goodwin:etal:2004b,Walch:etal:2010,Bate:etal:2010,Lomax:etal:2014,Fielding:etal:2015}.
Or systems may become misaligned in dense environments where encounters between multiple systems can destroy multiple systems \citep[e.g.][]{PKroupa:1995} or alter the separations, eccentricities and inclinations of companions \citep[e.g.][]{PKroupa:1995,Parker:Goodwin:2009}.
In particular, \citet{Parker:Goodwin:2009} find that $10-20$ per cent of binaries in an Orion Nebula-like cluster can be perturbed to inclination angles of $> 40\degr$ (the Kozai angle) by coplanar encounters\footnote{Note that there should be little or no effect from the Lidov-Kozai mechanism on the {\em disc} in a misaligned binary system.  The Lidov-Kozai timescale is given by $T \sim [M_{\primary}^{1/2}/M_{\secondary}][a^3/R^{3/2}](1 - e^2)^{3/2}$. For system with a primary mass $M_{\primary} = 0.5\Msun$, secondary mass $M_{\secondary} = 0.1\Msun$, semi-major axis $a = 300\AU$, disc radius $R = 50\AU$, and eccentricity $e = 0$, the timescale is $T \sim 0.5\mega\yr$. This timescale is much longer than that our disc uses in adjusting itself into a quasi-steady state, which is less than $1\kilo\yr$. Therefore any feature resulting from the mechanism in the disc would be erased.  However, planets formed in the misaligned would be subject to changes and perturbations from the Lidov-Kozai mechanism \cite[see][section 7]{Davies:etal:2014}.}.

In this paper we examine how discs in misaligned binary systems evolve.
In particular we investigate how the relative orientations of the disc, primary star and secondary star change as periodic star-disc interactions result in an exchange between the rotational angular momentum of the disc and the orbital angular momentum of the binary \citep[e.g.][]{Papaloizou:Terquem:1995,Bate:etal:2000}.

In section 2 we introduce the geometry and fundamentals of the problem.
In section 3 we describe the initial conditions of the simulations.
The results are then presented in section 4.
We analyse and discuss some of the results in section 5.
Finally, the conclusion is given in section 6.

\section{Precession and alignment in misaligned systems}\label{SECT:PRECESSION-AND-MISALIGNMENT-OF-MISALIGNED-SYSTEM}

In this section we present an analytic description of the precession and changes in alignment of the disc and companion in a misaligned binary system.
In summary we have a system with a primary at the centre and orbital planes associated with the disc and the companion.
The relative positions of these planes can be described by the angles between their respective angular momentum vectors.
When the disc and companion are misaligned, a torque is exerted on the disc by the companion (and vice versa) and the disc changes its alignment and also precesses.

\subsection{Coordinate systems}

At any one moment in time there are three coordinate systems that we are interested in.

{\bf Primary spin.}
The first system is defined by the rotation of the primary star with an angular momentum vector $\bm{J}_{\primary}$ which we assume is constant\footnote{The primary does accrete material from the disc, but this material is low angular momentum (otherwise it could not accrete), and has a negligible mass compared to the mass of the primary so this is a reasonable assumption.}.
This establishes an unchanging primary coordinate system ($x,y,z$) in which the primary is at (0,0,0) and $\bm{J}_{\primary}$ is in the positive $z$-direction.

{\bf Companion orbit.}
The second system $(x',y',z')$ is defined by the orbit of the companion star, also centred on the primary.
The companion has an instantaneous orbital angular momentum vector $\bm{J}_{\thebinary}$ in the positive $z'$-direction.
The $x'$- and $y'$-axes are chosen to be aligned with the semi-minor and semi-major axes of the companion's orbit respectively.
At any time the position of the companion can be described in polar coordinates $(r,\theta)$ on the $(x',y')$ plane.

More formally, the basis vectors for the companion coordinates $(x',y',z')$ are defined by $\hat{\bm{z}}' = \bm{J}_{\thebinary}/\vbrackets{\bm{J}_{\thebinary}}$, $\hat{\bm{x}}' = \hat{\bm{z}}\times\hat{\bm{z}}'/\vbrackets{\hat{\bm{z}}\times\hat{\bm{z}}'}$ and $\hat{\bm{y}}' = \hat{\bm{z}}'\times\hat{\bm{x}}'$.
Note that we fix the $(x',y',z')$ coordinate system so that it does not move.
Actually, the exchange of angular momentum between the disc and companion will alter the companion's orbit.
However, for the parameters we use the total angular momentum is dominated by the orbital angular momentum of the companion ($\vbrackets{\bm{J}_{\primary}} \ll \vbrackets{\bm{J}_{\disc}} \ll \vbrackets{\bm{J}_{\thebinary}}$) and so the companion's orbit does not change very much.

{\bf Disc rotation.}
The third system $(x'',y'',z'')$ is defined by the rotation of the disc (which is determined by an averaging process we describe later), and is again centred on the primary.
The disc angular momentum vector $\bm{J}_{\disc}$ defines the positive $z''$ direction, and the line of nodes (i.e. where the orbital plane of the binary crosses the rotational plane of the disc) defines $x''$.
Again, more formally, the basis vectors for $(x'',y'',z'')$ are $\hat{\bm{z}}'' = \bm{J}_{\disc}/\vbrackets{\bm{J}_{\disc}}$,  $\hat{\bm{x}}'' = \hat{\bm{z}}''\times\hat{\bm{z}}'/\vbrackets{\hat{\bm{z}}''\times\hat{\bm{z}}'}$ and $\hat{\bm{y}}'' = \hat{\bm{z}}''\times\hat{\bm{x}}''$.
Note that the orbits of the disc and binary can be prograde ($\bm{J}_{\disc}\cdot\bm{J}_{\thebinary} > 0$) or retrograde ($\bm{J}_{\disc}\cdot\bm{J}_{\thebinary} < 0$).

\begin{figure}
    \centering
    \includegraphics[angle=0,width=1.0\columnwidth]{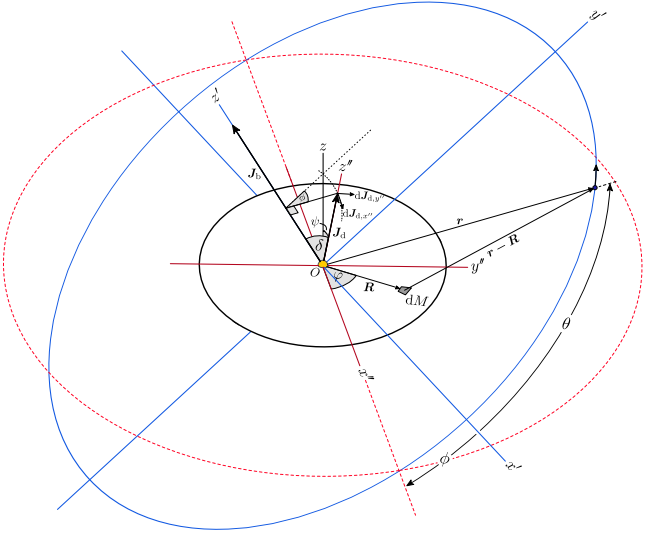}
    \caption{
        Illustration of a misaligned binary system.
        The system consists of (1) a primary star at the origin, (2) a circumprimary disc and (3) a secondary star orbiting at distance $r$ from the primary.
        The dashed red circle and the solid blue circle represent the rotational plane of the disc and the orbital plane of the binary respectively.
        The intersection between the two planes creates a line of nodes which is used as the $x''$-axis of the coordinate system $(x'',y'',z'')$.
        See text for the definitions of coordinates and variables.
    }
    \label{FIG:COMOVING-COORDINATES}
\end{figure}

The coordinate systems and angles are illustrated in Fig. \ref{FIG:COMOVING-COORDINATES}.
The primary is at the centre of the figure and the centre of the coordinate systems.
The companion's orbital plane is shown by the blue circle with the companion at a position $(r,\theta)$ (towards the top right).
The disc plane is shown by the dashed red circle.
In the disc plane is a mass element $\dee{M}$ at a distance $R$ from the primary, the importance of which we will describe later.

We then have three angles to consider:\\
1) The star-disc misalignment angle, $\psi$, between $\bm{J}_{\primary}$ and $\bm{J}_{\disc}$.\\
2) The companion-disc misalignment angle, $\delta$, between $\bm{J}_{\disc}$ and $\bm{J}_{\thebinary}$.\\
3) The companion-disc precession angle, $\phi$, which describes the precession of $\bm{J}_{\disc}$ about $\bm{J}_{\thebinary}$.

The initial values of $\psi$, $\delta$ and $\phi$ are denoted by $\psi_{\circ}$, $\delta_{\circ}$ and $\phi_{\circ}$ respectively.
The angles $\psi$, $\delta$ and $\phi$ are defined as follows.\\

\noindent(i) \textit{The star-disc misalignment angle} $\psi$, which is the angle between $\bm{J}_{\disc}$ and $\bm{J}_{\primary}$, is
\begin{equation}\label{EQ:DISC-PRIMARY-MISALIGNMENT-ANGLE-1}
    \psi = \cos^{-1}\rbrackets{\hat{\bm{z}}''\cdot\hat{\bm{z}}}.
\end{equation}\\

\noindent(ii) \textit{The disc-binary misalignment angle} $\delta$, which is the angle between $\bm{J}_{\disc}$ and $\bm{J}_{\thebinary}$, is
\begin{equation}\label{EQ:MISALIGNMENT-ANGLE-1}
    \delta = \cos^{-1}\rbrackets{\hat{\bm{z}}''\cdot\hat{\bm{z}}'}.
\end{equation}
The value of $\delta$ can also be used to determine whether the misaligned system is prograde ($0\degr \leq \delta < 90\degr$) or retrograde ($90\degr < \delta \leq 180\degr$).
There are two possible aligned systems -- when $\delta = 0\degr$, and when $\delta = 180\degr$.\\

\noindent(iii) \textit{The precession angle} $\phi$, that describes the precession of $\bm{J}_{\disc}$ about $\bm{J}_{\thebinary}$ or vice versa, is given by
\begin{equation}\label{EQ:PRECESSION-ANGLE-1}
    \phi =-\tan^{-1}\sbrackets{\frac{\hat{\bm{x}}''\cdot\hat{\bm{y}}'}{\hat{\bm{x}}''\cdot\hat{\bm{x}}'}}.
\end{equation}
The precession angle is positive in a prograde system and negative in a retrograde system.\\

It is worth noting that the angle $\psi$ may also be written in terms of $\phi$ and $\delta$.
In a simple case where the change in the angle $\delta$ is small, $\bm{J}_{\disc}$ will precess almost circularly about $\bm{J}_{\thebinary}$.
The chord that subtends the angle $\psi$ (with a unit length of $2\sin(\psi/2)$) is approximately equal to that which subtends the angle $\phi$ (with a unit length of $2\sin\delta|\sin(\phi/2)|$).  We then have
\begin{equation}\label{EQ:DISC-PRIMARY-MISALIGNMENT-ANGLE-2}
    \psi \simeq 2\sin^{-1}\sbrackets{\sin\delta_{\circ}\vbrackets{\sin\rbrackets{\frac{\phi}{2}}}},
\end{equation}
where $\delta_{\circ}$ is the initial misalignment angle between the disc and the binary.

\begin{figure}
    \centering
    \includegraphics[angle=270,width=1.0\columnwidth]{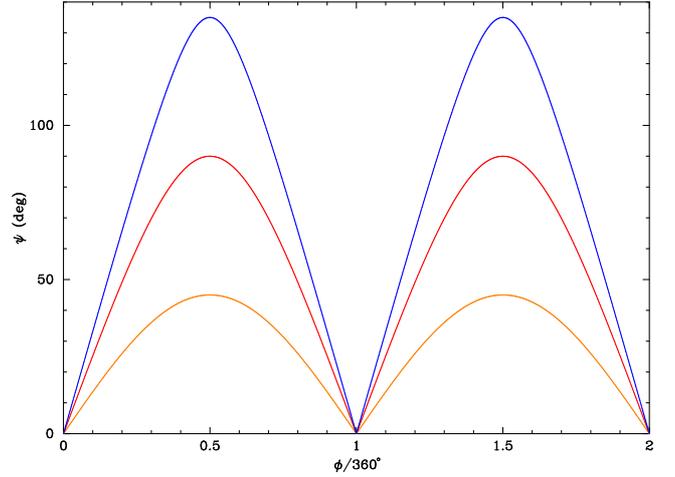}
    \caption{
        The star-disc misalignment angle $\psi$ as a function of the precession angle $\phi$ given in equation \eqref{EQ:DISC-PRIMARY-MISALIGNMENT-ANGLE-2}.
        Orange, red and blue lines are of systems with initial misalignment angle $\delta_{\circ} = 22.5\degr$, $45\degr$ and $67.5\degr$ respectively.
        The disc in the $67.5\degr$-misaligned system can be considered as temporarily retrograde when $\psi > 90\degr$.
    }
    \label{FIG:ANGLE-PSI}
\end{figure}

As the disc precesses ($\phi$ changes from zero through $360\degr$), the star-disc misalignment angle ($\psi$) changes.
Figure \ref{FIG:ANGLE-PSI} shows the change in the star-disc misalignment angle $\psi$ with the precession angle $\phi$ for (initial) disc-binary misalignment angles $\delta_{\circ} = 22.5\degr$, $45\degr$ and $67.5\degr$.
The value of $\psi$ oscillates between $0\degr$ and $2\delta_{\circ}$.
For example, when $\delta_{\circ} = 22.5\degr$, $\psi$ varies between $0\degr$ and $45\degr$; and when $\delta_{\circ} = 67.5\degr$, $\psi$ varies between $0\degr$ and $135\degr$.
In a system with $\delta_{\circ} > 45\degr$, the rotation of the disc could thus be temporarily retrograde ($\psi > 90\degr$) with respect to the spin axis of the primary star.

\subsection{Rigid-disc approximation}\label{SECT:RIGID-DISC-APPROXIMATION}

In this subsection we outline the rigid-disc approximation \citep[e.g.][]{Korycansky:Papaloizou:1995}.
In misaligned systems, the disc tilts as a consequence of the net torque $\bm{T}'' = (T_{x''},T_{y''},T_{z''})$ exerted on the disc.
The rates of change of the angles $\phi$ and $\delta$ are related to the torque and the angular momentum of the disc.
For precession, the precession rate $\dot{\phi}$ is associated with the component torque $T_{x''}$ exerted on the disc along the $x''$-axis.
An infinitesimal change in $\bm{J}_{\disc}$ along the $x''$-axis due to  $T_{x''}$ can be written as $\dee{J_{\disc,x''}} = J_{\disc}\sin\delta\dee{\phi}$ (see Fig. \ref{FIG:COMOVING-COORDINATES}).
Since $\dee{J_{\disc,x''}}/\dee{t} = T_{x''}$, the precession rate can then be written as
\begin{equation}\label{EQ:PHI-DOT-1}
    \dot{\phi} = \frac{T_{x''}}{J_{\disc}}\csc\delta.
\end{equation}
In similar manner, the (disc-binary) alignment rate $\dot{\delta}$, which is associated with $T_{y''}$ exerted along the $y''$-axis, can be written as
\begin{equation}\label{EQ:DELTA-DOT-1}
    \dot{\delta} = \frac{T_{y''}}{J_{\disc}}.
\end{equation}

Here we show how the torque $\bm{T}''$ on the disc from the companion is determined.
We simplify the problem by considering the disc as a rigid body.
The bulk motion of the disc will depend only on \textit{the gravitational torque} exerted on the disc as a single object.
We call this kind of torque \textit{the rigid-body torque} to distinguish it from \textit{the tidal torque}, which involves the fluidity of the disc.
This approximation clearly ignores the tidal torque and other kinds of torque such as \textit{the encounter torque} \citep[e.g.][for coplanar systems]{Korycansky:Papaloizou:1995}.

The rigid-body torque on the rigid disc is
\begin{align}\label{EQ:NET-TORQUE-1}
    \bm{T} = \int_{\mathrm{disc}}\bm{R} \times \bm{f} \,\, \dee{M},
\end{align}
where $\bm{f}$ is the force per unit mass (acceleration), due to the secondary star, exerted on a mass element $\dee{M}$ at radius $R$ on the disc (see Fig. \ref{FIG:COMOVING-COORDINATES}).
For a flat disc with surface density $\Sigma = \Sigma(R)$, the mass element is given by $\dee{M} = \Sigma{}R\dee{\varphi}\dee{R}$, where $\varphi$ is the azimuthal angle of the disc.

From Fig. \ref{FIG:COMOVING-COORDINATES}, the force per unit mass exerted on the mass element $\dee{M}$ can be written as
\begin{equation}\label{EQ:FORCE-PER-UNIT-MASS-1}
    \bm{f} = GM_{\secondary}\frac{\bm{r}-\bm{R}}{\vbrackets{\bm{r}-\bm{R}}^{3}},
\end{equation}
where $G$ is the gravitational constant and $\bm{r}$ is the position vector of the secondary star of mass $M_{\secondary}$.
The magnitude of $\bm{r}$ is
\begin{equation}\label{EQ:MAGNITUDE-OF-rb-1}
    r = \frac{a(1-e^{2})}{1-e\sin\theta},
\end{equation}
where $a$ and $e$ are the orbital semi-major axis and the orbital eccentricity of the companion.
The term $1/\vbrackets{\bm{r}-\bm{R}}^{3}$ in equation \eqref{EQ:FORCE-PER-UNIT-MASS-1} can be written as $1/r^{3}(1+\varepsilon)^{3/2}$, where $\varepsilon = \sbrackets{R^{2}-2(\bm{R}\cdot\bm{r})}/r^{2}$.
For a disc with radius $R \ll r$, the alternative form of the term can be expanded by Taylor series.
To the first-order approximation, one can find that (after substituting and rearranging)
\begin{equation}\label{EQ:FORCE-PER-UNIT-MASS-2}
    \frac{1}{\vbrackets{\bm{r}-\bm{R}}^{3}} \simeq \frac{1}{r^{3}}-\frac{3}{2}\frac{R^{2}}{r^{5}}+3\frac{{\bm{R} \cdot \bm{r}}}{r^{5}}.
\end{equation}
The integrand of equation \eqref{EQ:NET-TORQUE-1} thus becomes
\begin{equation}\label{EQ:R-X-f-1}
    \bm{R} \times \bm{f} \simeq \frac{GM_{\secondary}}{r^{3}}\sbrackets{1-\frac{3}{2}\frac{R^{2}}{r^{2}}+3\frac{{\bm{R} \cdot \bm{r}}}{r^{2}}}\rbrackets{{\bm{R} \times \bm{r}}}.
\end{equation}
In order to integrate equation \eqref{EQ:NET-TORQUE-1} to obtain the component torques for equation \eqref{EQ:PHI-DOT-1} and \eqref{EQ:DELTA-DOT-1}, it is convenient to use the disc coordinates to describe the position vectors $\bm{R}$ and $\bm{r}$.

In terms of the coordinates $(x'',y'',z'')$, the position vector $\bm{R}$ of the mass element $\dee{M}$ is
\begin{equation}\label{EQ:VECTOR-R}
    \bm{R} = R\bigl(\cos\varphi,\,\sin\varphi,\,0\bigr)
\end{equation}
and the position vector $\bm{r}$ of the secondary star is
\begin{equation}\label{EQ:VECTOR-rb}
    \bm{r} = r\bigl(\cos(\theta+\phi),\,\sin(\theta+\phi)\cos\delta,\,\sin(\theta+\phi)\sin\delta\bigr).
\end{equation}
One can find that the products of the vectors $\bm{R}$ and $\bm{r}$ are
\begin{equation}\label{EQ:R-DOT-rb-1}
    \bm{R} \cdot \bm{r} = Rr\sbrackets{\cos(\theta+\phi)\cos\varphi+\sin(\theta+\phi)\cos\delta\sin\varphi}
\end{equation}
and
\begin{align}\label{EQ:R-X-rb-1}
    \begin{split}
        \bm{R} \times \bm{r} = Rr\bigl(&\sin(\theta+\phi)\sin\delta\sin\varphi,\,\\
                                              -&\sin(\theta+\phi)\sin\delta\cos\varphi,\\
                                               &\sin(\theta+\phi)\cos\delta\cos\varphi-\cos(\theta+\phi)\sin\varphi\bigr).
    \end{split}
\end{align}
By substituting equation \eqref{EQ:R-DOT-rb-1} and (\ref{EQ:R-X-rb-1}) into equation \eqref{EQ:R-X-f-1}, we have $\bm{R} \times \bm{f}$ required to solve equation \eqref{EQ:NET-TORQUE-1}.
Integrating equation \eqref{EQ:NET-TORQUE-1} over the entire disc, i.e. $\varphi$ from $0$ to $2\pi$ and $R$ from $0$ to $R_{\disc}$, gives us a net instantaneous torque exerted on the rigid disc.
Using the fact that only terms with `$\sin^{2}\varphi$' or `$\cos^{2}\varphi$' can survive from the integration over the given range of $\varphi$, we finally have
\begin{align}\label{EQ:TORQUE-ON-ENTIRE-DISC-1}
    \begin{split}
        \bm{T}'' \simeq   &\frac{3\pi}{2}\frac{GM_{\secondary}}{r^{3}}\rbrackets{\int_{0}^{R_{\disc}}\Sigma{}R^{3}\dee{R}}\\
                            &{\bigl(2{\sin^{2}}{(\theta+\phi)}\sin\delta\cos\delta,\,-\sin[2(\theta+\phi)]\sin\delta,\,0\bigr)},
    \end{split}
\end{align}
where $\bm{T}'' = (T_{x''},T_{y''},T_{z''})$.
In the disc coordinates, the component $T_{z''}$ is zero because gravitational forces acting along the disc midplane cancel out by the symmetry of the rigid disc.

The integral term in equation \eqref{EQ:TORQUE-ON-ENTIRE-DISC-1} can be calculated by adopting a power-law surface density of index $p$, i.e. $\Sigma = \Sigma_{\circ}(R/R_{\circ})^{-p}$.
One finds that
\begin{equation}
    \int_{0}^{R_{\disc}}\Sigma{}R^{3}\dee{R} = \frac{\Sigma_{\circ}R_{\circ}^{p}R_{\disc}^{4-p}}{4-p}.
\end{equation}

\subsection{Precession and alignment rates}\label{SECT:PRECESSION-AND-ALIGNMENT-RATES}

The angular momentum $J_{\disc}$ required for equation \eqref{EQ:PHI-DOT-1} and \eqref{EQ:DELTA-DOT-1} for a flat disc can be obtained from considering an annulus of radius $R$, width $\dee{R}$ and tangential velocity $v = \rbrackets{GM_{\primary}/R}^{1/2}$, where $M_{\primary}$ is the mass of the primary star.
The angular momentum of an annulus is $\dee{J}_{\disc} \simeq Rv\dee{M} = 2\pi\Sigma\rbrackets{GM_{\primary}}^{1/2}R^{3/2}\dee{R}$.
Integrating from $R = 0$ to $R_{\disc}$ gives
\begin{equation}
    J_{\disc} \simeq \frac{4\pi\Sigma_{\circ}R_{\circ}^{p}\rbrackets{GM_{\primary}}^{1/2}}{5-2p}R_{\disc}^{5/2-p}.
\end{equation}
By substituting this $J_{\disc}$ and the associated components of torque from equation \eqref{EQ:TORQUE-ON-ENTIRE-DISC-1} into equation \eqref{EQ:PHI-DOT-1} and \eqref{EQ:DELTA-DOT-1}, we find that the instantaneous precession rate is
\begin{equation}\label{EQ:PHI-DOT-2}
    \dot{\phi} \simeq  2\eta\frac{R_{\disc}^{3/2}}{r^{3}}\sin^{2}(\theta+\phi)\cos\delta
\end{equation}
and the instantaneous alignment rate is
\begin{equation}\label{EQ:DELTA-DOT-2}
    \dot{\delta} \simeq -\eta\frac{R_{\disc}^{3/2}}{r^{3}}\sin[2(\theta+\phi)]\sin\delta,
\end{equation}
where
\begin{equation}
    \eta = \frac{3}{8}\rbrackets{\frac{5-2p}{4-p}}\rbrackets{\frac{GM_{\secondary}^{2}}{M_{\primary}}}^{1/2}.
\end{equation}

In practice, it is more convenient to use the time-averaged forms of equation \eqref{EQ:PHI-DOT-2} and \eqref{EQ:DELTA-DOT-2} than the instantaneous forms.
The time-averaged precession rate can be found from averaging equation \eqref{EQ:PHI-DOT-2} over one orbital period $P$, i.e.
\begin{equation}\label{EQ:TIME-AVERAGED-PHI-1}
    \dot{\abrackets{\phi}} = \frac{1}{P}\int_{0}^{P}\dot{\phi}\,\dee{t} = \frac{1}{P}\int_{0}^{2\pi}\dot{\phi}\,\frac{\dee{\theta}}{\dot{\theta}}.
\end{equation}
By substituting $\dot{\theta} = 2\pi{}a^{2}\sqrt{1-e^{2}}/r^{2}P$ and assuming that the changes in other variables are negligible compared to $\theta$, we have
\begin{equation}\label{EQ:TIME-AVERAGED-PHI-2}
    \dot{\abrackets{\phi}} \simeq \eta\sbrackets{\frac{R_{\disc}}{a^{2}(1-e^{2})}}^{3/2}\cos\delta.
\end{equation}
This precession rate is essentially the same as that obtained in, for example, \citet{Bate:etal:2000}.

Similarly, one can find that the time-averaged alignment rate can be written as
\begin{equation}\label{EQ:TIME-AVERAGED-DELTA-1}
    \dot{\abrackets{\delta}} \simeq -\frac{\dot{\abrackets{\phi}}\tan\delta}{2\pi}\int_{0}^{2\pi}\sin\sbrackets{2\rbrackets{\theta+\phi}}(1-e\sin\theta)\dee{\theta}.
\end{equation}
However, if we neglect the change in $\phi$ as we do in finding $\dot{\abrackets{\phi}}$, the integral is zero (i.e. no net change in the alignment of the rigid disc).
However, the angle $\phi$ does change over an orbital period, and thus the net change is non-zero.
We will discuss this later in Section \ref{SECT:ANALYSIS-ALIGNMENT-RATES}.

\section{Simulation set-up}\label{SECT:SIMULATION-SET-UP}

In this work, we perform smooth particle hydrodynamic \citep[SPH;][]{Gingold:Monaghan:1977,Lucy:1977,Monaghan:1992} simulations to investigate the bulk evolution of circumprimary discs in misaligned systems.
Simulations are performed by using the high-performance SPH code \textsc{seren} \citep{Hubber:etal:2011}.
The code use the method introduced by \citet{Stamatellos:etal:2007} to treat radiative heating and cooling in the disc.

The procedures begins with creating a star-disc system whose main star is represented by a sink particle \citep[see e.g.][]{Bate:etal:1995} of primary mass $M_{\primary} = 0.5\Msun$ and accretion radius $0.5\AU$.
The disc has initial mass $M_{\disc} = 0.07\Msun$, inner radius $R_{\mathrm{in}} = 0.5\AU$, and outer radius $R_{\mathrm{out}} = 40\AU$.
This  system is evolved for $1\kilo\yr$ to ensure that the disc is in a quasi-steady state.

We then create a misaligned binary system by adding a sink particle to represent the secondary star of mass $M_{\secondary} = 0.1\Msun$ and accretion radius $0.5\AU$ with various semi-major axes, eccentricities and initial inclinations.

We assume that this represents a physical situation in which a binary system with a wide companion has formed a circumprimary disc in isolation from the (distant) companion.
An encounter then perturbs the orbit of the companion causing it to begin interacting with the disc.

\subsection{Isolated star-disc systems}

Here we present the initial conditions of the primary-disc system that is relaxed before adding the companion.
The method of constructing an SPH disc can be found in Appendix A.

\subsubsection{Density and temperature profiles}

The initial disc has a power-law function for the initial surface density
\begin{equation}\label{EQ:SURFACE-DENSITY-PROFILE-1}
    \Sigma(R) = \Sigma_{1}\rbrackets{\frac{R}{1\AU}}^{-p},
\end{equation}
where $p = 0.5$ is the power-law index and $\Sigma_{1}$ is the surface density at radius $R = 1\AU$.

The value of $\Sigma_{1}$ can be calculated by supposing that the disc is flat, so that the mass of an annular strip of radius $R$ can be written as $\dee{M} = 2\pi\Sigma R\dee{R}$.
Comparing the integrated mass of the disc in radius $0.5 \leq R \leq
40\AU$ with $M_{\disc} = 0.07\Msun$ gives us $\Sigma_{1} =
6.615\times10^{-5}\Msun\,\AU^{-2}$ ($\sim 588 \mathrm{g\,cm^{-2}}$) for our disc.
We note that the initial values of $p$ and $\Sigma_{1}$ are not crucial, since particles in the disc will quickly be redistributed according to the (artificial and real) viscosity and temperature structure of the disc (see the results below).

For the temperature structure of the disc, we use a modified power-law function
\begin{equation}\label{EQ:TEMPERATURE-PROFILE-1}
    T(R) = T_{1}\rbrackets{\frac{R}{1\AU}}^{-q}+T_{\infty},
\end{equation}
where $q$ is the power-law index, $T_{1}$ the temperature at $R = 1\AU$, and $T_{\infty} = 10\Kelvin$ the background radiation temperature.

Unlike density, the underlying {\em minimum} temperature is imposed on particles in the disc depending on their {\em current} distance from the primary.
That is, particles at radius $R$ will have temperature at least $T(R)$ (it may be higher due to shocking but is not allowed to fall below this value).

In this work, we test the stability of discs with temperature indices $q = 0.5$ (flared disc), $0.75$ (flat disc) and $1$; and temperatures $T_{1} = 300\Kelvin$, $600\Kelvin$ and $1200\Kelvin$.

For the main set of simulations, systems have a disc with index $q = 0.75$, temperature $T_{1} = 300\Kelvin$ (see Section \ref{SECT:ISOLATED-STAR-DISC-SYSTEMS}), and resolution after relaxing (see below) slightly less than $300\kilo$ particles.

\subsubsection{Viscosity and resolution}

The artificial viscosity parameters $\alpha_{\rm\scriptscriptstyle{SPH}}$ and $\beta_{\rm\scriptscriptstyle{SPH}}$ have a major role in controlling the artificial viscosity in SPH simulations \citep[e.g.][]{Monaghan:1997,Price:2008}.
Our simulations use the standard viscosity prescription with $\alpha_{\rm\scriptscriptstyle{SPH}} = 0.1$ and $\beta_{\rm\scriptscriptstyle{SPH}} = 0.2$.
We do not use any additions to viscosity (e.g. the Balsara viscosity switch \citep{Balsara:1995}).

Most simulations are performed with an initial resolution of $3\times10^{5}$ particles.
A number of simulations with lower resolution ($1.5\times10^{5}$ particles) and higher resolution ($6\times10^{5}$ particles) are also performed to investigate convergence.

Simulations are terminated when the disc is represented by less than $10^{4}$ particles as at this point we cannot resolve any moderately realistic disc structure and cannot believe the evolution of the disc at all.

\subsection{Constructing misaligned binary systems}

After allowing our disc to relax for $1\kilo\yr$ we add a (misaligned) companion star of mass $M_{\secondary} = 0.1\Msun$.
The companion star is launched at the apastron radius $r_{\textrm{max}} = 300\AU$ from the primary.
The orbital configurations of the binary are selected from the combinations of (1) initial misalignment angles $\delta_{\circ} = 22.5\degr-157.5\degr$, in steps of $22.5\degr$, and (2) initial eccentricities $e_{\circ} = 0-0.6$, in steps of $0.2$.
Note that systems with $\delta_{\circ} > 90\degr$ are retrograde.

\subsection{Determining the rotational axis of the disc}

In misaligned systems, we expect to see precession and a change in the alignment between the disc and the primary and the disc and the companion (as measured by their angular momentum vectors
$\bm{J}_{\disc}$, $\bm{J}_{\thebinary}$ and $\bm{J}_{\primary}$).
These are described in terms of the angles $\phi$, $\delta$ and $\psi$.

In order to calculate these angles, we must firstly define the basis vectors of the coordinates from $\bm{J}_{\disc}$ and $\bm{J}_{\thebinary}$.
However, the vector $\bm{J}_{\disc}$ cannot be uniquely determined in the simulation, because the extent of the disc is not well-defined after being perturbed by the secondary star and distortions may appear in the disc.
Fortunately, we only require the direction of $\bm{J}_{\disc}$ (not its magnitude) and we calculate this from the average angular momentum vector of SPH particles within $40\AU$ from the primary.
We find that this range contains both a sufficient number of particles to avoid noise when the resolution is low, and is close enough to not include particles belonging to any secondary disc that may form in highly eccentric systems (the closest separation between the stars is $75\AU$ for companions with $e = 0.6$).

\section{Results}\label{SECT:RESULTS}

\subsection{Stability of isolated star-disc systems}\label{SECT:ISOLATED-STAR-DISC-SYSTEMS}

Before proceeding to examine the evolution of circumprimary discs in misaligned binary systems, we will first investigate the evolution of a circumprimary disc in isolation.
In this subsection, we justify our choice of discs with the parameters $q=0.75$ and $T_{1}=300\Kelvin$ as being long-lived and stable in isolation and being reasonable representations of real discs.

\begin{figure}
    \centering
    \includegraphics[angle=270,width=0.99\columnwidth]{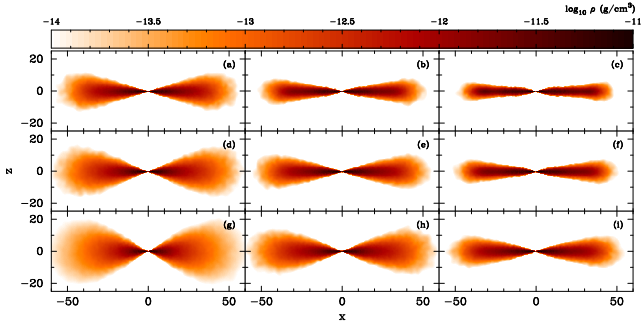}
    \caption{
        Cross-sectional density plots of star-disc systems after evolved for $1\kilo\yr$.
        Each system has different temperature structures, parameterized by the index $q$ and the temperature $T_{1}$.
        Panels in the same column have the same index $q$: from left to right, $q = 0.5$, $0.75$, and $1$.
        Panels in the same row have the same temperature $T_{1}$: from top to bottom, $T_{1} = 300\Kelvin$, $600\Kelvin$, and $1200\Kelvin$.
    }
    \label{FIG:DISC-CROSS-SECTIONS}
\end{figure}

In the first few hundred years, the disc rapidly adjusts itself into a quasi-steady state, where quantities such as the density at a given position change gradually with time.
With radiative heating and cooling treated by the \citet{Stamatellos:etal:2007} method, the vertical structure of the disc is characterised by its temperature structure, as shown in Fig. \ref{FIG:DISC-CROSS-SECTIONS}.
This figure shows cross-sectional density plots for discs with different temperatures ($300\Kelvin$, $600\Kelvin$ and $1200\Kelvin$ from the top row to the bottom row), and temperature indices ($q = 0.5$, $0.75$ and $1$ from the left column to the right column).
Discs with both lower-$q$ and higher-$T_{1}$ (towards the bottom left) are `fluffier' as they have higher temperatures at a given radius.
We also find that the hotter the disc, the higher the accretion rate onto the central star as pressures throughout the disc pushing particles into the inner regions and the (empty) sink.

We typically relax discs for $1\kilo\yr$ before adding a companion star.
However, to test the long-term stability we have evolved the discs further until the resolution is less than $50\kilo$ particles.
We wish to find disc parameters that produce long-lived and gravitationally stable discs to use as our initial discs.
The reason for this is that we do not want secular  processes to drive disc evolution, rather we wish to ensure that the changes to the disc are driven by the companion.

The gravitational stability of the disc at radius $R$ from the central star can be expressed in terms of the Toomre parameter $Q(R)$ \citep{Toomre:1964} and the cooling time parameter $\beta_{\mathrm{cool}}(R)$ \citep{Gammie:2001}.
The disc is considered to be gravitationally \textit{unstable} if $Q(R) \lesssim 1$ \textit{and} $\beta_{\mathrm{cool}}(R) \lesssim 3$.
If only one of the criteria is met, however, the disc is still gravitationally stable.

\begin{figure}
    \centering
    \includegraphics[angle=270,width=0.99\columnwidth]{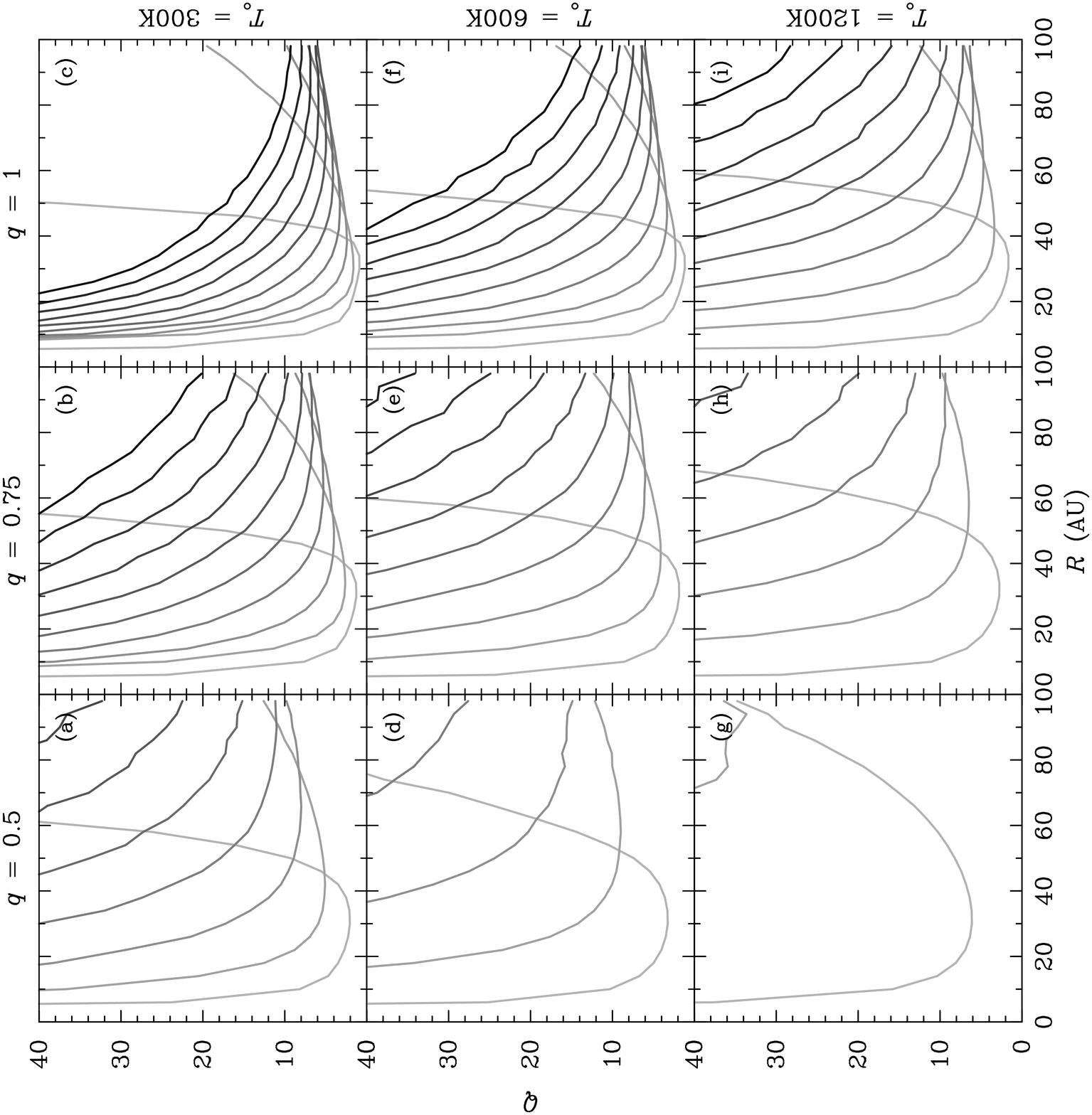}\\ \vspace{0.1in}
    \includegraphics[angle=270,width=0.99\columnwidth]{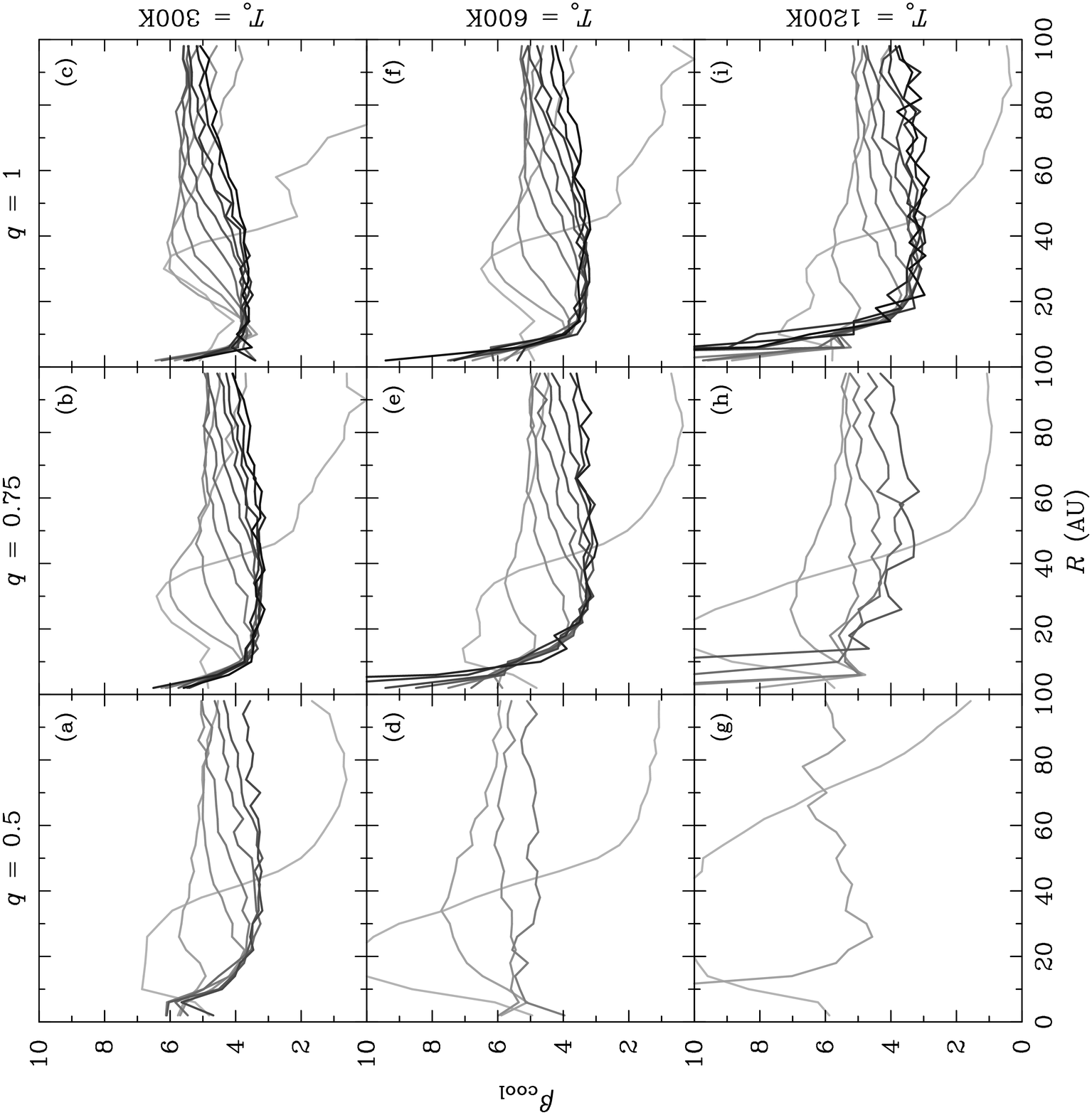}
    \caption{
        Changes in Toomre parameter $Q$ (top panels) and cooling time parameter $\beta_{\mathrm{cool}}$ (bottom panels) of discs with various temperature structures shown in Fig. \ref{FIG:DISC-CROSS-SECTIONS}.
        Lines in each panel are of snapshots at times from $t = 1\kilo\yr$ (lightest grey) in steps of $50\kilo\yr$ (darker grey).
    }
    \label{FIG:Q-AND-BETACOOL}
\end{figure}

The values of $Q(R)$ and $\beta_{\mathrm{cool}}(R)$ of the discs in Fig. \ref{FIG:DISC-CROSS-SECTIONS} from time $t = 1\kilo\yr$ (lightest grey) in steps of $50\kilo\yr$ (darker grey) until the resolution is less than $50\kilo$ particles are shown in Fig. \ref{FIG:Q-AND-BETACOOL}.
We see that all discs at $t > 1\kilo\yr$ have $Q(R)$ and/or $\beta_{\mathrm{cool}}(R)$ well above the instability criteria.
Therefore all discs are gravitationally stable and unlikely to fragment spontaneously.

\begin{figure}
    \centering
    \includegraphics[angle=270,width=0.99\columnwidth]{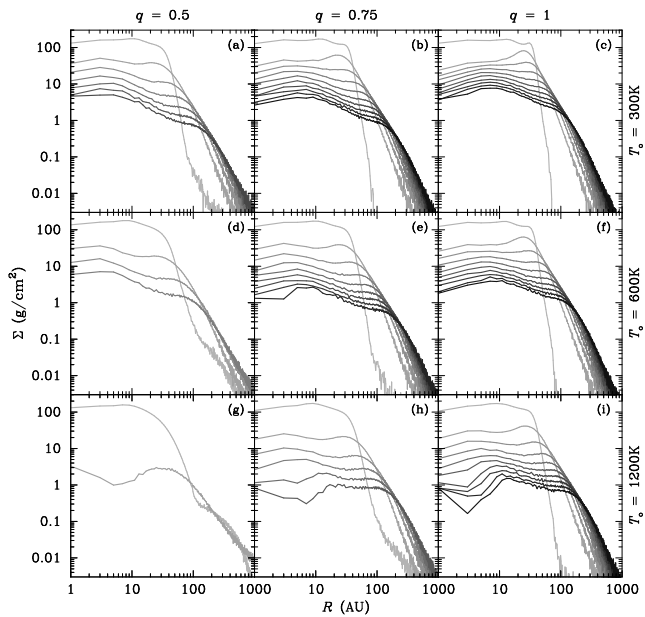}
    \caption{
        Changes in density profiles (surface density, $\Sigma$, against distance, $R$) of discs in the same set as Fig. \ref{FIG:Q-AND-BETACOOL}, with the same line colouring.
    }
    \label{FIG:SIGMA_VS_R}
\end{figure}

We also require that our discs be long-lived as well as gravitationally stable.
Figure \ref{FIG:SIGMA_VS_R} shows the density profiles of the discs from Fig. \ref{FIG:DISC-CROSS-SECTIONS} in steps of $50\kilo\yr$, as in Fig. \ref{FIG:Q-AND-BETACOOL}.
The more lines present in Fig. \ref{FIG:SIGMA_VS_R} the longer the disc lives (more $50\kilo\yr$ snapshots present) and the closer together the lines are, the less the disc has changed its density profile with time.
We see that colder discs (higher-$q$ and lower-$T_{1}$ towards the top right) live longer and change less than hotter discs.
This is unsurprising in our simulations as the higher pressure in the hotter discs drives accretion onto the primary and then the disc must readjust.

In terms of both stability and lifetime, the coldest disc with $q = 1$ and $T_{1} = 300\Kelvin$ (top right panel in all plots so far) would be the best choice for the simulation.
However, observation and theory suggest the value of the power-law index $q \sim 0.43-0.75$ and the temperature $T_{1} \sim 100-400\Kelvin$ \citep[e.g.][]{Pringle:1981,Chiang:Goldreich:1997,Andrews:Williams:2007}.
Therefore we use discs with $q = 0.75$ and $T_{1} = 300\Kelvin$ (top middle panel in all plots so far) as a compromise between numerical/physical stability and longevity and as a match to real discs.

\subsection{The evolution of discs in binary systems}\label{SECT:EVOLUTION-OF-DISCS-IN-BINARY-SYSTEMS}

In this subsection, we present the results of our investigation of discs in misaligned binary systems.
We label simulations with a name of the form e.g. `\texttt{pe6d450}' which contains the information on the companion orbit. \\
(1) The first character is the orbital direction, \texttt{p} for prograde and \texttt{r} for retrograde.\\
(2) The initial eccentricity $e_{\circ}$, e.g. \texttt{e2} for $e_{\circ} = 0.2$. \\
(3) The initial misalignment angle $\delta_{\circ}$, e.g. \texttt{d675} for $\delta_{\circ} = 67.5\degr$.\\
So in our example above of `\texttt{pe6d450}' this is a prograde companion orbit with an eccentricity of $0.6$ and an initial misalignment angle of $45\degr$.
Initial conditions and labels of the main set of simulations are listed in Table \ref{TAB:INITIAL-CONDITIONS}.
Later we will also introduce the initial number of particles in the disc, but for now all simulations have $300\kilo$ particles.

Note that the results throughout depend on resolution, and in particular on the relationship between the SPH artificial viscosity and the effective disc viscosity that this gives which is highly non-trivial \citep[see e.g.][]{Lodato:Price:2010,Rosotti:etal:2014}.
Therefore care should be taken in comparing only the general trends of our results with analytic predictions: exact numbers/timescales etc. will almost certainly vary depending on the resolution, exact form of the artificial viscosity etc. and should be treated with caution.

\begin{table}
    \caption{Orbital configurations and labels for the main set of simulations.  $\delta_{\circ}$ is the initial companion misalignment angle, $e_{\circ}$ is the initial companion eccentricity, and `Label' is the shorthand used to refer to the simulations in the text and figures.}\label{TAB:INITIAL-CONDITIONS}
    \begin{tabular}{ccc}
    \hline\hline
    $\delta_{\circ}$ & $e_{\circ}$ & Label\\
    \hline
    $22.5\degr$                 & $0$       & \texttt{pe0d225}\\
    \hline
    \multirow{4}{*}{$45\degr$}  & $0$       & \texttt{pe0d450}\\
                                & $0.2$     & \texttt{pe2d450}\\
                                & $0.4$     & \texttt{pe4d450}\\
                                & $0.6$     & \texttt{pe6d450}\\
    \hline
    $67.5\degr$                 & $0$       & \texttt{pe0d675}\\
    \hline
    $90\degr$                   & $0$       & \texttt{pe0d900}\\
    \hline
    $112.5\degr$                & $0$       & \texttt{re0d675}\\
    \hline
    \multirow{2}{*}{$135\degr$} & $0$       & \texttt{re0d450}\\
                                & $0.4$     & \texttt{re4d450}\\
    \hline
    $157.5\degr$                & $0$       & \texttt{re0d225}\\
    \hline
    \end{tabular}
\end{table}

\subsubsection{Aligned systems}

Before discussing misaligned systems it is worth very quickly considering aligned (coplanar) systems.
We find that tidal perturbations from the companion have a negligible effect on the stability of the disc, and the disc remains stable through to the end of the simulation.
The outward transfer of angular momentum in the disc makes the disc expand and fill its Roche lobe on the plane, resulting in mass transfer from the disc to the companion star.
The transferred mass forms a secondary disc around the companion which rotates in the same direction as that of the primary (counterclockwise).

In aligned systems with highly elliptical orbits ($e_{\circ} = 0.6$), the tidal effects of the companion are strong enough to generate a spiral density waves in the disc during the pericentric passage.
The mass accretion rate of the primary star, however, changes only slightly as a response to the passage of the companion, even in the system with $e_{\circ} = 0.6$.
Therefore the tidal effects of our binary companions are not very significantly affecting the stability or longevity of our discs.
This allows us to examine misaligned discs with some confidence that
effects we see are largely down to their misalignment.

\subsubsection{Precession in misaligned systems}

The angular momentum vector of the disc will precess about the angular momentum vector of the companion, and the angle at any time is $\phi$.

\begin{figure}
    \centering
    \includegraphics[angle=270,width=0.99\columnwidth]{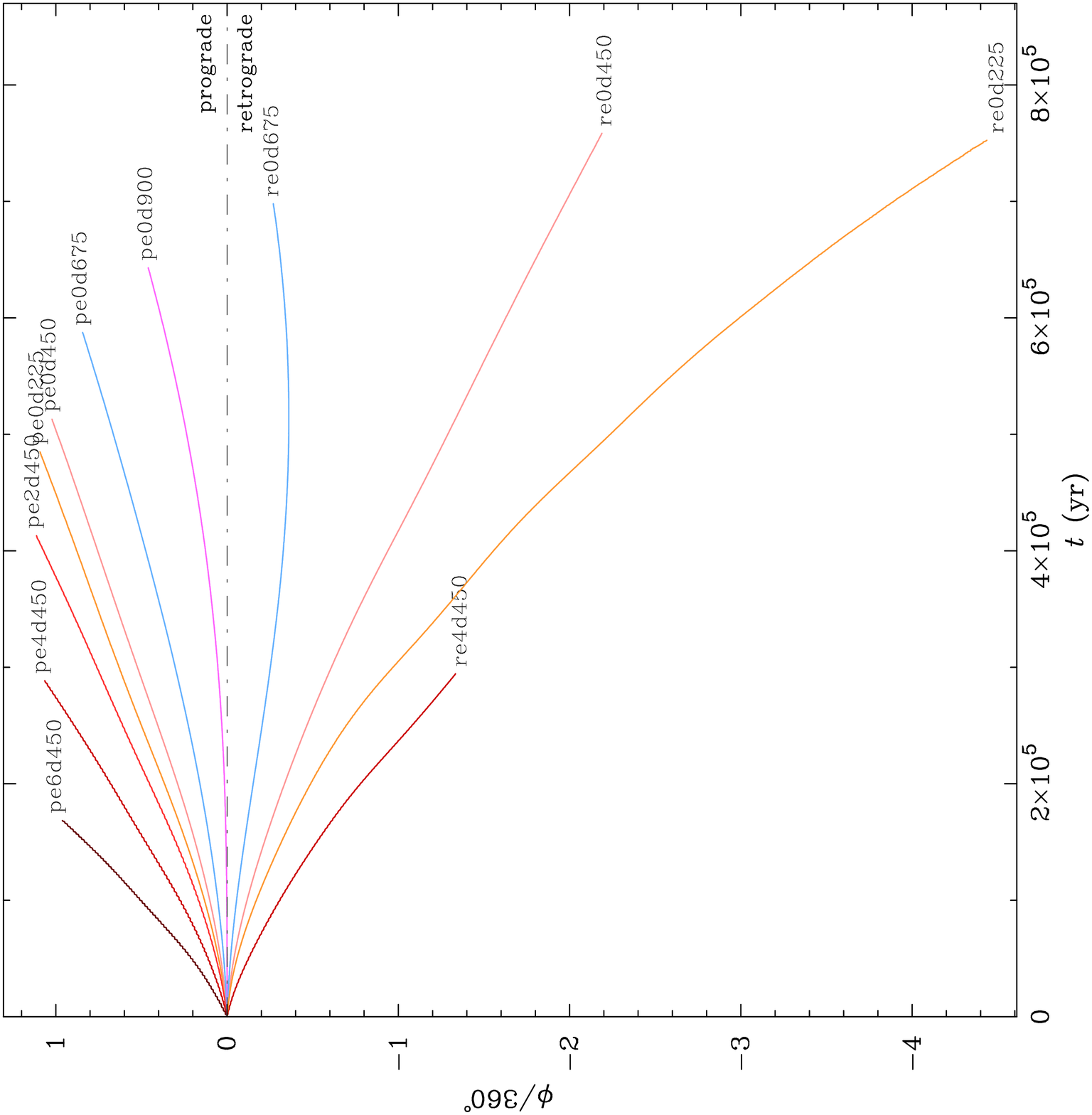}
    \caption{
        Changes in the precession angle ($\phi$) against time ($t$) for systems with different initial misalignment angles and eccentricities (for labels see text).
    }
    \label{FIG:PHI-VS-T-GENERAL}
\end{figure}

The change in the precession angle $\phi$ with time of systems in the main set of simulations are shown in Fig. \ref{FIG:PHI-VS-T-GENERAL}.
On the $y$-axis is the precession angle $\phi$ in multiples of $360\degr$ evolving with time (on the $x$-axis) for less than $800\kilo\yr$.
A positive change in $\phi$ is prograde precession, and negative change is retrograde precession.
For example, discs that precess $\phi/360\degr = -4$ have completed $4$ retrograde cycles.
Each curve ends at the time when the disc resolution is less than $10\kilo$ particles (i.e. we can no longer believe our results at all).

The first (unsurprising) thing to note in Fig. \ref{FIG:PHI-VS-T-GENERAL} is that prograde misaligned companions produce prograde precession (positive $\phi$), and retrograde companions produce retrograde precession (negative $\phi$).

The next thing to note is that the speed of precession (or how fast the angle $\phi$ changes) depends on the initial misalignment angle $\delta_{\circ}$ and eccentricity $e_{\circ}$ of the companion.

For systems with the same eccentricity, those with $\delta_{\circ}$ closer to either $0\degr$ or $180\degr$ (coplanar) precess faster.
For example (see Fig. \ref{FIG:PHI-VS-T-GENERAL}), in the prograde system, \texttt{pe0d225} (orange line above $\phi=0$) precesses faster than \texttt{pe0d450} (light red line above $\phi=0$) and \texttt{pe0d675} (light blue line above $\phi=0$).
Similarly, in retrograde system, \texttt{re0d225} (orange line below $\phi=0$) precesses faster than \texttt{re0d450} (light red line below $\phi=0$) and \texttt{re0d675} (light blue line below $\phi=0$).

For systems with the same misalignment angle, higher eccentric systems, with either prograde or retrograde orbit, precess faster.
In particular compare those prograde systems with $\delta_{\circ} = 45\degr$ (four red lines above $\phi = 0$ in Fig. \ref{FIG:PHI-VS-T-GENERAL}), \texttt{pe6d450} precesses faster than \texttt{pe4d450}, \texttt{pe2d450} and \texttt{pe0d450} (darker line is steeper).

This behaviour is consistent with that predicted by equation equation \eqref{EQ:TIME-AVERAGED-PHI-2} where the time-averaged precession rate goes as $\cos\delta$ and $(1-e^2)^{-3/2}$.
We investigate this later in Section \ref{SECT:ANALYSIS-PRECESSION-RATES}.

\subsubsection{Disc-companion misalignment angle}\label{SECT:THE-MISALIGNMENT-ANGLES}

\begin{figure}
    \centering
    \includegraphics[angle=270,width=0.99\columnwidth]{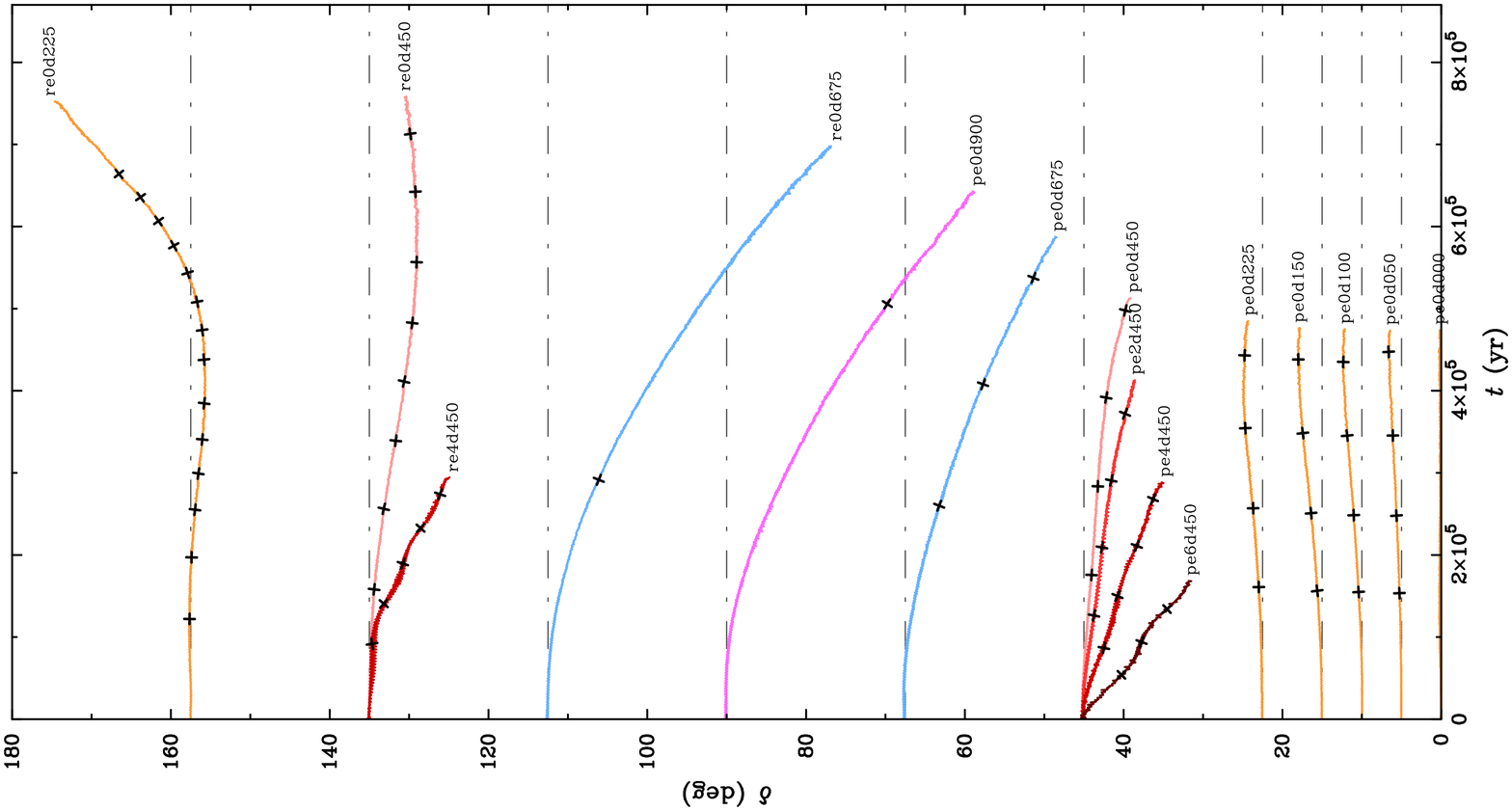}
    \caption{
        Changes in the misalignment angle $\delta$ in prograde and retrograde misaligned systems with time ($t$).
        Cross symbols are marked on each curve, at times when $\phi$ is an integer multiple of $90\degr$, to indicate how fast the precession is compared to the change in $\delta$ or vice versa.
        For the meaning of the labels see text.
    }
    \label{FIG:DELTA-VS-T-GENERAL}
\end{figure}

As well as precessing, the disc-companion misalignment angle $\delta$ changes from its initial value of $\delta_{\circ}$, as a consequence of angular momentum transfers between the objects.
The change in $\delta$ with time for various values of $\delta_{\circ}$ (the value of $\delta$ at $t = 0$) is shown in Fig. \ref{FIG:DELTA-VS-T-GENERAL}.
This includes a number of systems with low values of $\delta_{\circ}$ ($15\degr$, $10\degr$, $5\degr$ and $0\degr$ towards the bottom of the plot not illustrated in Fig. \ref{FIG:PHI-VS-T-GENERAL}).
Note that in all cases, the rate of change in alignment is much lower than the rate of precession.

In Fig. \ref{FIG:DELTA-VS-T-GENERAL} we can see that systems with $\delta_{\circ} \geq 45\degr$ tend to adjust themselves towards the $0\degr$-alignment, with the alignment rate ($\dot{\delta}$) depending on the misalignment angle $\delta$ and the eccentricity $e$.
In the four systems with $\delta_{\circ} = 45\degr$ towards the bottom of Fig. \ref{FIG:DELTA-VS-T-GENERAL}, the fastest realignment is for the $e = 0.6$ system (\texttt{pe6d450} towards the left), and the slowest for the $e = 0$ system (\texttt{pe0d450} towards the right).
Moving up Fig. \ref{FIG:DELTA-VS-T-GENERAL} we see that systems with $\delta_{\circ}$ close to $90\degr$ (e.g. \texttt{pe0d675}, \texttt{pe0d900} and \texttt{re0d675}) attempt to align themselves more rapidly.
The retrograde system \texttt{re0d675} can even bring itself to be prograde with $\delta$ changing from $112.5\degr$ to $\sim 70\degr$.
However, this trend becomes reverse in systems with $\delta_{\circ}$ close to $0\degr$ or $180\degr$, e.g. \texttt{pe0d050-225}, \texttt{re0d225} and \texttt{re0d450}: the systems attempt to misalign themselves instead.
This behaviour is crucial as it can prevent misaligned systems from being aligned.
{\em Misaligned systems may not be able to align themselves within the disc lifetime.}
The behaviour is due to the presence of two torques, which we will return to in the next subsection.

In addition, another characteristic of change in alignment is the nodding motion between the disc midplane and the orbital plane.
The motion makes the value of $\delta$ oscillate two times per binary orbit, as suggested by the term $\sin\sbrackets{2\rbrackets{\theta+\phi}}$ in equation \eqref{EQ:DELTA-DOT-2}.
The amplitude of the oscillation is only a fraction of degree, just enough to make the curves in Fig. \ref{FIG:DELTA-VS-T-GENERAL} look slightly irregular.

\subsubsection{Changes in the companion orbit}\label{SECT:CHANGES-IN-BINARY-ORBITS}

\begin{figure}
    \centering
    \includegraphics[angle=270,width=0.99\columnwidth]{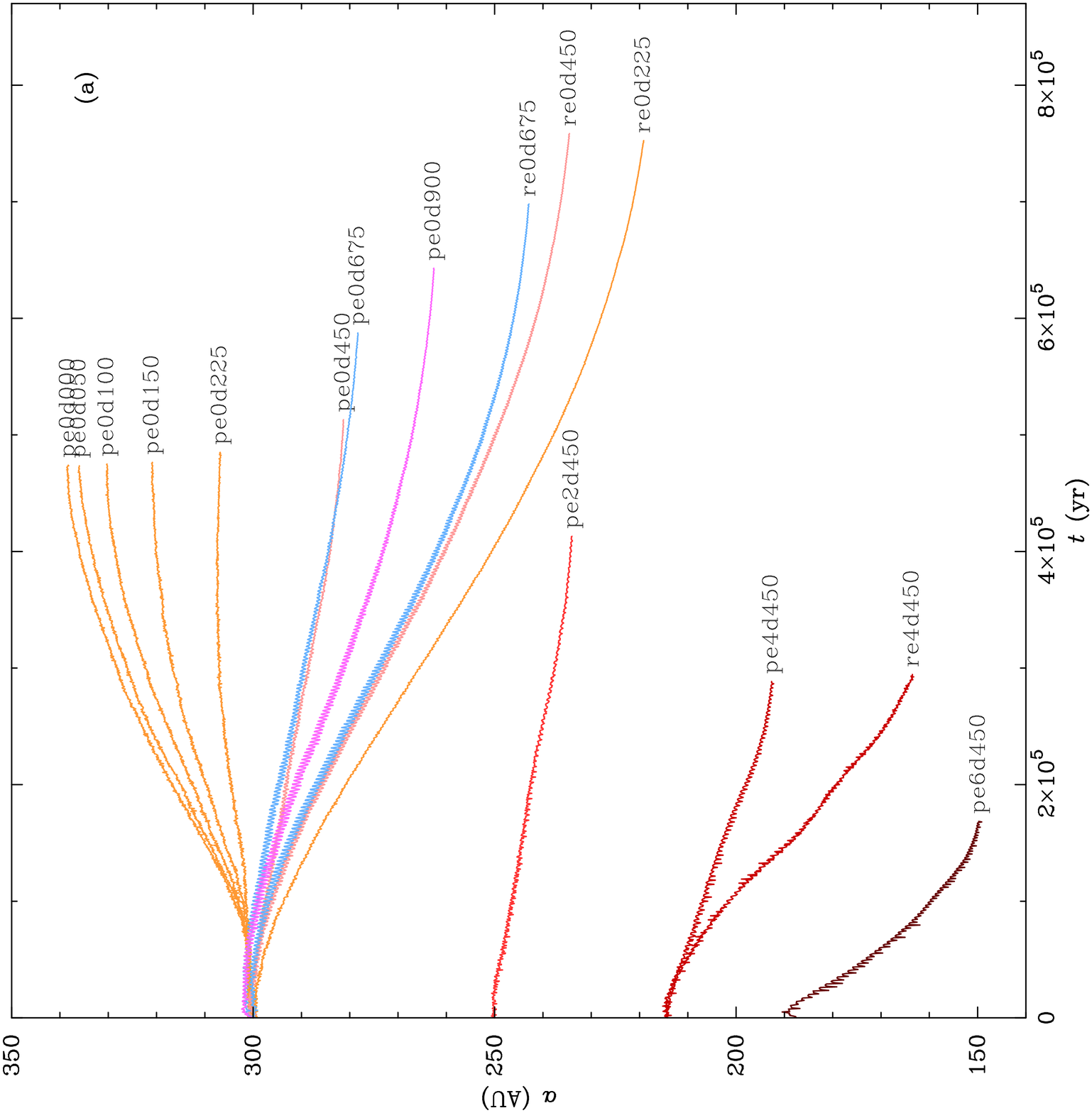}\\\vspace{0.1in}
    \includegraphics[angle=270,width=0.99\columnwidth]{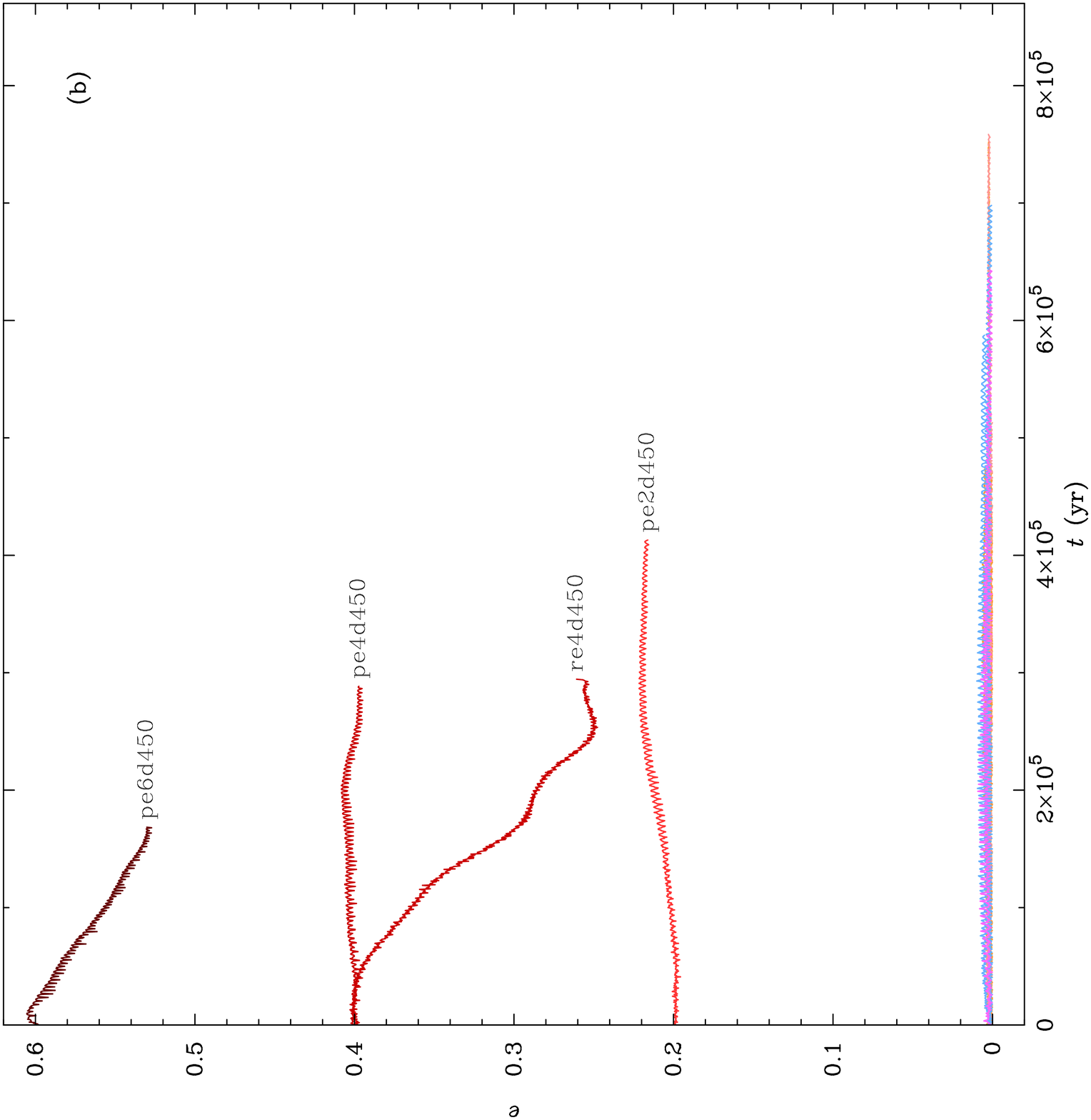}
    \caption{
        Changes in the semi-major axis, $a$, (top panel), and eccentricity, $e$, (bottom panel) with time, $t$.
        Systems with $e = 0$ are not labelled in panel (b) as their curves change insignificantly and lay on top of one another.
        See text for the meaning of the labels.
    }
    \label{FIG:A-E-VS-T-GENERAL}
\end{figure}

As well as the companion changing the orientation of the disc, the disc causes changes in the semi-major axis ($a$), the eccentricity ($e$), and the orientation of the companion.
The orientation of the binary is given by the orbital angular momentum $\bm{J}_{\thebinary}$ which is also related to $a$ and $e$ via $J_{\thebinary} \propto \sqrt{a(1-e^{2})}$, where $J_{\thebinary}$ is the magnitude of $\bm{J}_{\thebinary}$.
Changes in $a$ and $e$ of all the systems in Fig. \ref{FIG:DELTA-VS-T-GENERAL} are shown in Fig. \ref{FIG:A-E-VS-T-GENERAL}.

The top panel of Fig. \ref{FIG:A-E-VS-T-GENERAL} shows the evolution of the companion's semi-major axis.
All systems have the same initial apastron of $300\AU$ but different semi-major axes as they have different eccentricities.
The bottom panel of Fig. \ref{FIG:A-E-VS-T-GENERAL} shows the evolution of the eccentricities for $e = 0$, $0.2$, $0.4$ and $0.6$.
Note that all $e = 0$ systems are laid on top of one another at the bottom of the plot, as zero-eccentricity orbits do not change their eccentricities significantly.

What is clear from both panels is that the change in $a$ and $e$ is not simple.
In particular, $a$ can either increase or decrease in different systems.
Here, changes in the orbital parameters tell us about the net torque exerted on the binary orbit.

If we consider the change in $a$ in a prograde system with $e = 0$ (circular orbit) then the magnitude of $\bm{J}_{\thebinary}$ is only proportional to $\sqrt{a}$.
From Fig. \ref{FIG:A-E-VS-T-GENERAL}(a), the change in $a$ tell us that $J_{\thebinary}$ tends to increase in systems with low $\delta$ (e.g. \texttt{pe0d225-000}) and decrease in systems with high $\delta$ (e.g. \texttt{pe0d450-900}).
In terms of the torque, this implies that the component of torque exerted on the binary orbit changes its direction from parallel to $\bm{J}_{\thebinary}$, in low-$\delta$ systems, to anti-parallel to $\bm{J}_{\thebinary}$, in high-$\delta$ systems, at some value of $\delta$ (between $22.5\degr$ and $45\degr$).

This can be understood if we consider the torque as the sum of two (or more) torques acting against each other and having amplitudes which vary with $\delta$.
The two most plausible torques are the tidal torque and the encounter torque.

The tidal torque is due to the gravitational interaction between the companion and the \textit{fluid} disc.

The encounter torque is due to a drag force exerted during encounters between the companion and the disc.
Since the direction of the drag force acting on the companion is mainly against the azimuthal direction of motion, the direction of the encounter torque would be more or less opposite to that of $\bm{J}_{\thebinary}$.
Hence, the encounter torque always tends to decrease the magnitude of $\bm{J}_{\thebinary}$.
From Fig. \ref{FIG:A-E-VS-T-GENERAL}(a), the influence of the encounter torque seems to increase with a decrease in the periastron (compare \texttt{pe0d450} with \texttt{pe4d450}) and an increase in the companion-disc relative velocity (compare \texttt{pe0d450} with \texttt{re0d450}).

The direction of the tidal torque, on the other hand, can be determined from the change in $J_{\thebinary}$ (or $a$) in prograde systems with low $\delta$, where the tidal interaction dominates the encounter interaction (as the encounter velocity is close to zero).
The increase in $J_{\thebinary}$ in systems from \texttt{pe0d225} to \texttt{pe0d000} suggests that the direction of the tidal torque is roughly the same as that of $\bm{J}_{\thebinary}$, i.e. the torque tends to increase the magnitude of $\bm{J}_{\thebinary}$.
In systems with $\delta > 0\degr$, the direction of the tidal torque would lie somewhere between $\bm{J}_{\thebinary}$ and $\bm{J}_{\disc}$.
We will discuss the roles of the torques in changing $\delta$ in Section \ref{SECT:ANALYSIS-NON-RIGID-DISC-MODEL} below.

\subsubsection{Star-disc misalignment angle}\label{SECT:STAR-DISC-MISALIGNMENT-ANGLE}

\begin{figure}
    \centering
    \includegraphics[angle=270,width=0.99\columnwidth]{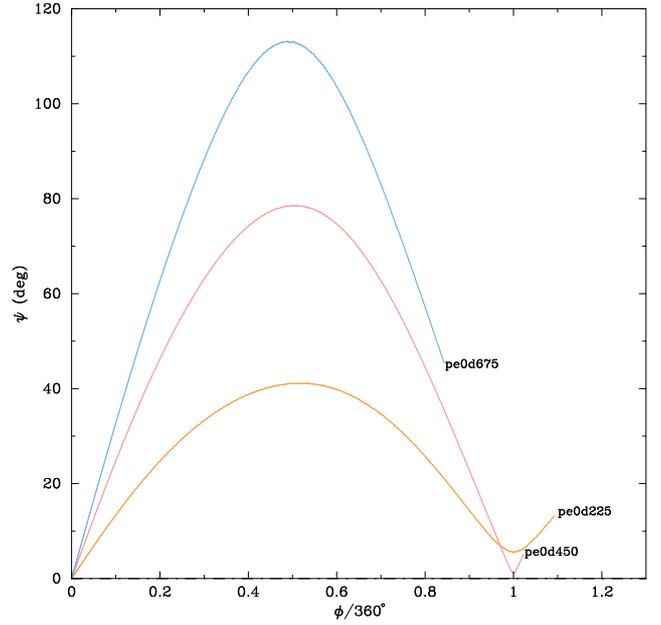}
    \caption{
        Changes in the star-disc misalignment angle $\psi$ in systems \texttt{pe0d225}, \texttt{pe0d450} and \texttt{pe0d675}.
        Compare this figure to Fig. \ref{FIG:ANGLE-PSI}, where $\delta$ is assumed to be constant.
    }
    \label{FIG:PSI-VS-T}
\end{figure}

The star-disc misalignment angle ($\psi$) changes periodically as a consequence of precession. Figure \ref{FIG:PSI-VS-T} shows how the angle $\psi$ changes with respect to $\phi$ in systems \texttt{pe0d225}, \texttt{pe0d450} and \texttt{pe0d675}, i.e. the same systems as shown in Fig. \ref{FIG:ANGLE-PSI}.
We can see that the value of $\psi$ varies between $0$ and $< 2\delta_{\circ}$, as suggested by equation \eqref{EQ:DISC-PRIMARY-MISALIGNMENT-ANGLE-2}.

The change in the angle $\psi$ due to precession is one possible explanation for the spin-orbit misalignment found in exoplanetary systems \citep[][]{Batygin:2012}.
Since the disc midplane defines the orbital plane of planets that form in the disc, the orbital axis of the planets would later be misaligned from the spin axis of the central star if the original star-disc system has $\psi > 0$.

\subsubsection{Resolution and numerics}

\begin{figure}
    \centering
    \includegraphics[angle=270,width=0.99\columnwidth]{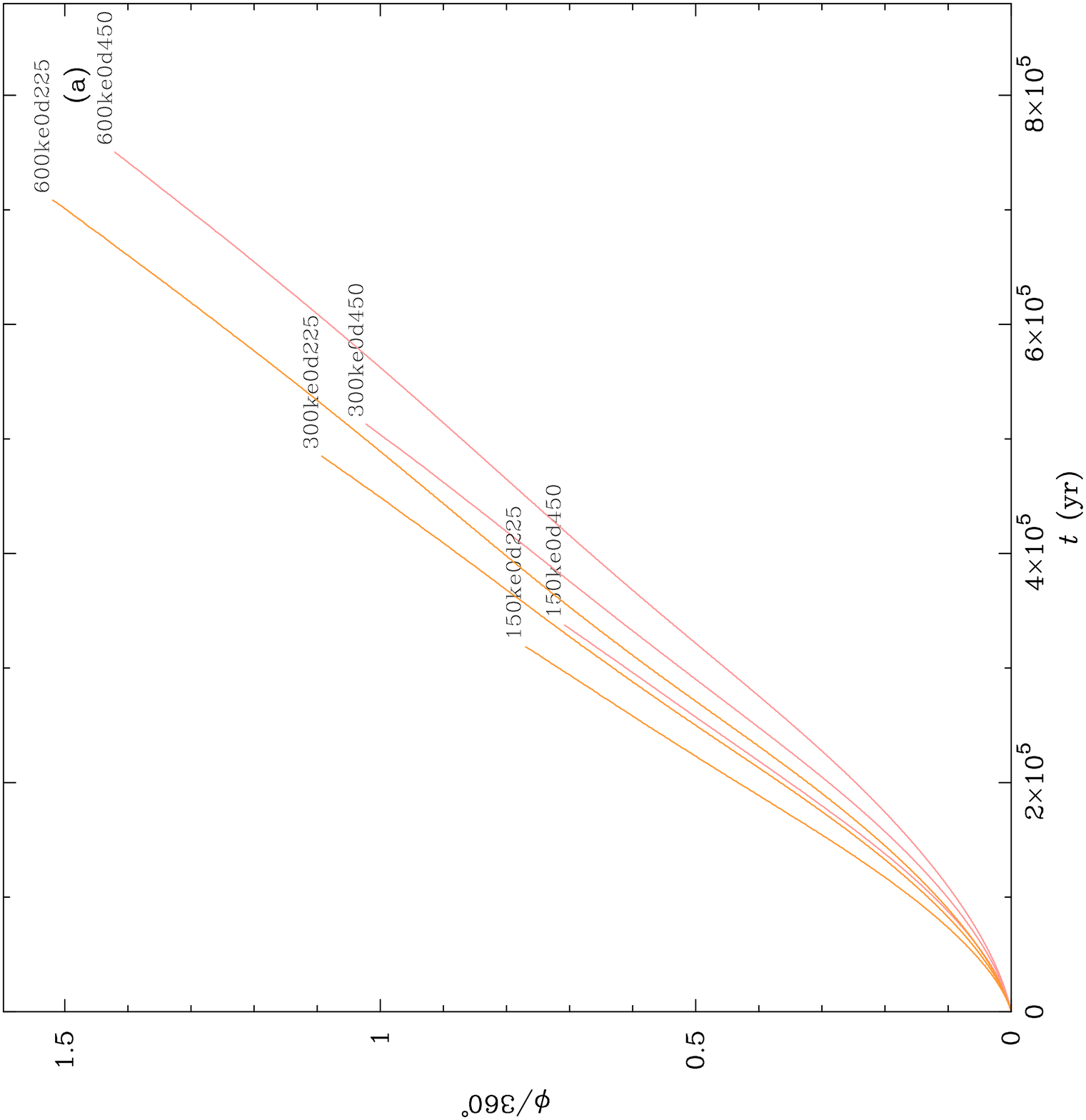}\vspace{0.1in}
    \includegraphics[angle=270,width=0.99\columnwidth]{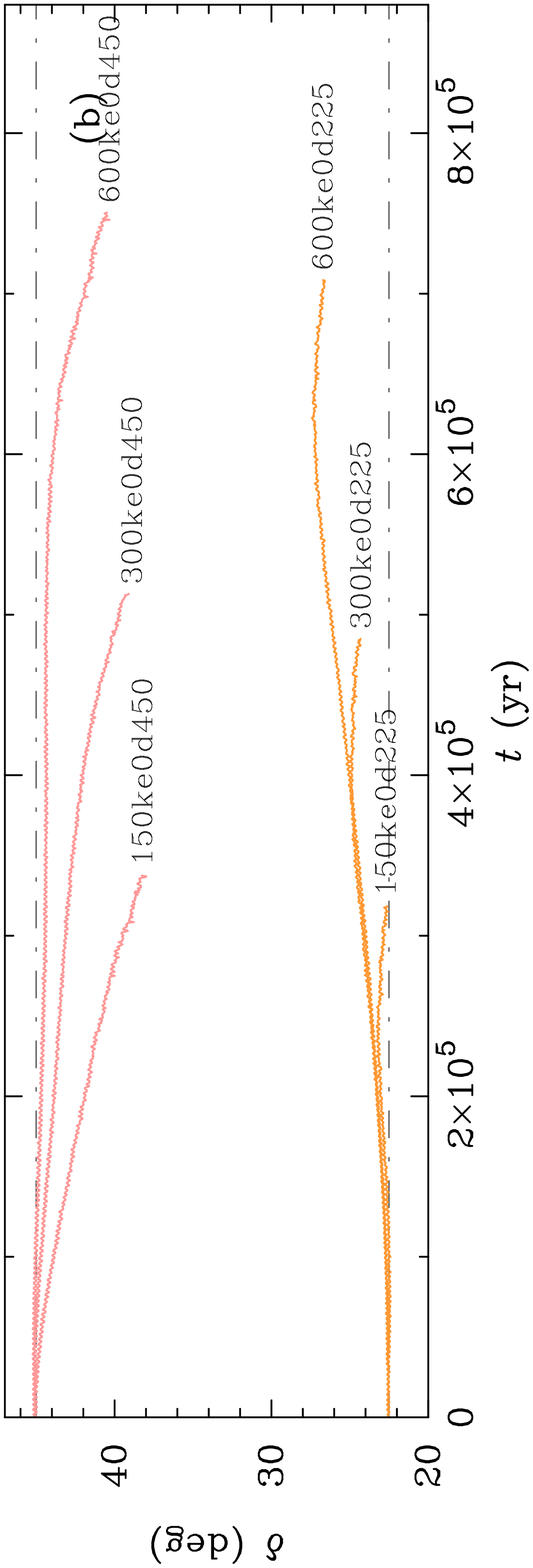}
    \caption{
        Changes in the precession angle ($\phi$, top panel), and the misalignment angle ($\delta$, bottom panel) with time ($t$).   For systems \texttt{pe0d225} and \texttt{pe0d450} with discs of $150\kilo$, $300\kilo$ and $600\kilo$ particles.
    }
    \label{FIG:EFFECTS-OF-RESOLUTION}
\end{figure}

Resolution is one of the major numerical issues in disc simulations using SPH.
In our simulated misaligned systems, using lower resolution (particle numbers) tends to increase the rates of precession and change in alignment, as shown in Fig. \ref{FIG:EFFECTS-OF-RESOLUTION}.

The top panel of Fig. \ref{FIG:EFFECTS-OF-RESOLUTION} shows the change in the precession angle $\phi$ with time for different resolutions for $150\kilo$, $300\kilo$, and $600\kilo$ particles in the initial disc.
The particle number proceeds the usual simulation code, e.g. \texttt{150ke0d225} and \texttt{600ke0d225} are simulations with zero eccentricity and an initial misalignment of $22.5\degr$ but with $150\kilo$ and $600\kilo$ particles respectively.
For both $22.5\degr$ (orange lines) and $45\degr$ (red lines) initial misalignments, $\phi$ changes more slowly at higher resolutions.

The lower panel of Fig. \ref{FIG:EFFECTS-OF-RESOLUTION} shows the change in alignment angle $\delta$ with time (same legend as the panel above).
Again, we see that higher resolution discs change their alignments more slowly.

The lower rates of change when higher resolution is used are due to the decreased effect of artificial viscosity.
However, we see that both high and low resolution discs have very similar final states, despite the higher resolution discs taking a longer (physical) time to reach this state.

\section{Analysis and discussion}\label{SECT:ANALYSIS}

\subsection{Precession rates}\label{SECT:ANALYSIS-PRECESSION-RATES}

The aim of this subsection is to examine the consistency of the rigid-disc model in describing the precession process.
By examining the evolution of the precession angles in Fig. \ref{FIG:PHI-VS-T-GENERAL}, we see that the precession rates are roughly consistent with the analytic derivation in equation \eqref{EQ:TIME-AVERAGED-PHI-2}.
That is, systems that precess rapidly have low $\delta$ and high $e$ (or small $a$).

We now compare the precession rate calculated from equation \eqref{EQ:TIME-AVERAGED-PHI-2} with that found in the simulations (by means of polynomial regression).
In calculating $\dot{\abrackets{\phi}}$ from equation \eqref{EQ:TIME-AVERAGED-PHI-2}, the values of all the input parameters are taken directly from the result, except that of the power-law index $p$ and the radius $R_{\disc}$ that are obtained from estimation.

The value of the index $p$ is the slope of the surface density curve within a certain radial range (see Fig. \ref{FIG:SIGMA_VS_R}).
To avoid noise and complications from potentially low-density and disturbed outer regions, we calculate the value of $p$ only within $40\AU$ from the central star.
As the system evolves and the disc changes, the value of $p$ changes.
In most cases, the value varies between $-0.14$ and $0.33$.
However, the value of $p$ does not significantly affect the calculation of $\dot{\abrackets{\phi}}$.

The radius $R_{\disc}$, on the other hand, is the important parameter in equation \eqref{EQ:TIME-AVERAGED-PHI-2} since $\dot{\abrackets{\phi}}$ is proportional to $R_{\disc}^{3/2}$.  Unfortunately, $R_{\disc}$ is not a well-defined parameter.
We determine $R_{\disc}$ using the weighted average radius
\begin{equation}\label{EQ:WEIGHT-AVERAGE-R}
    R_{\mathrm{av}} = \frac{\int_{0}^{M_{\disc}}R\dee{M}}{\int_{0}^{M_{\disc}}\dee{M}} = \frac{\int_{0}^{R_{\mathrm{max}}}\Sigma{R^{2}}\dee{R}}{\int_{0}^{R_{\mathrm{max}}}\Sigma{R}\dee{R}},
\end{equation}
where $\dee{M} = 2\pi\Sigma R\dee{R}$ and $R_{\mathrm{max}}$ is the extent of the disc.

\begin{figure}
    \centering
    \includegraphics[angle=270,width=0.99\columnwidth]{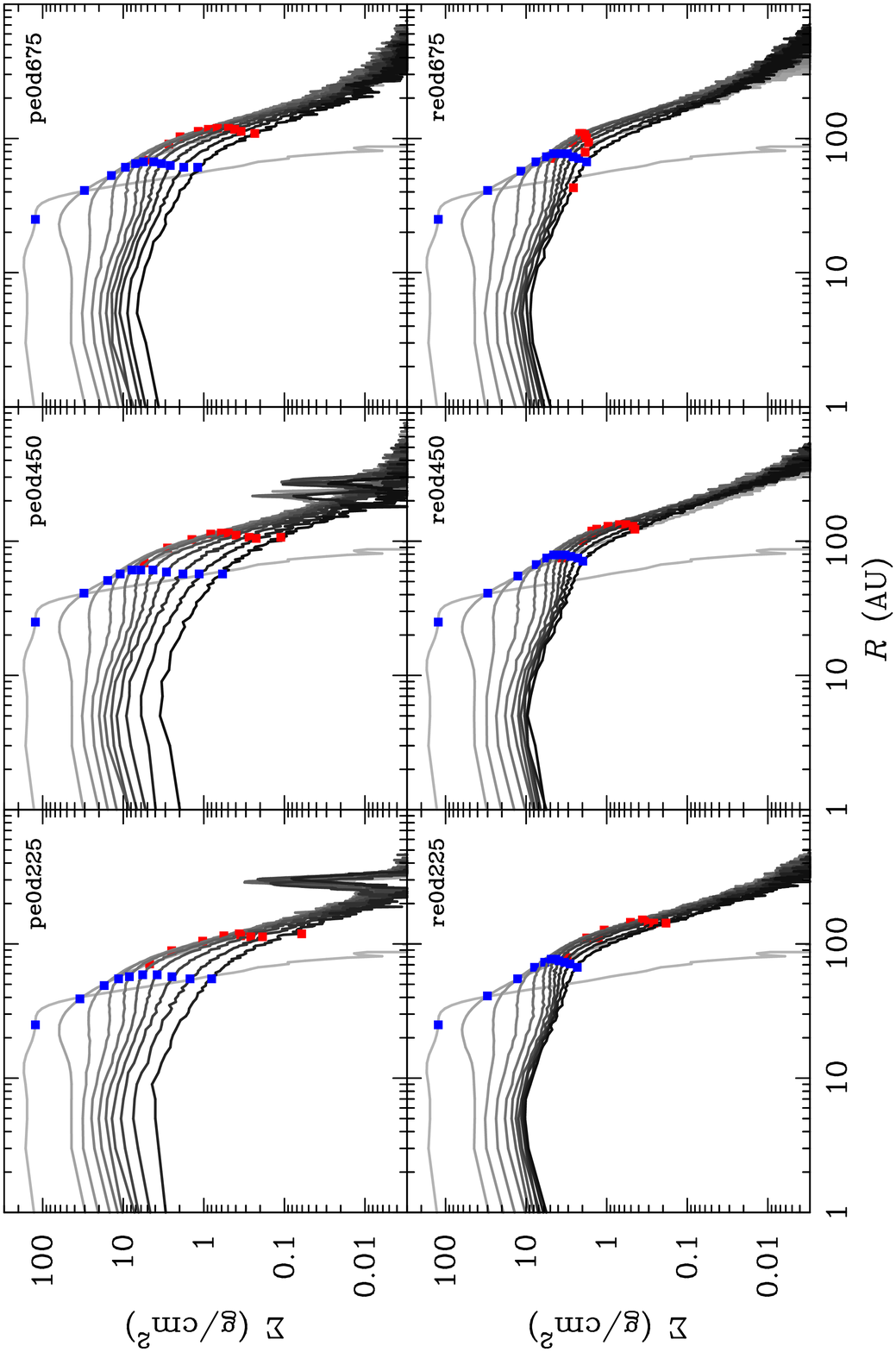}
    \caption{
        Density profiles of discs in the selected systems at various times from $t = 0$ (lightest grey) progressing in steps of $50\kilo\yr$ (darker grey).
        Small peaks on the profiles at $R \sim 200-300\AU$, clearly seen in systems \texttt{pe0d225} and \texttt{pe0d450}, are of the disc surrounding the companion (the secondary disc).
        Blue square symbols mark values of $R_{\mathrm{av}}$, which are calculated from equation \eqref{EQ:WEIGHT-AVERAGE-R}.
        Red square symbols mark values of $R_{\mathrm{eff}}$ (see text).
    }
    \label{FIG:SIGMA-VS-R-GENERAL}
\end{figure}

Let us consider systems with zero-eccentricity orbits such as \texttt{pe0d225}, \texttt{pe0d450}, \texttt{pe0d675} and their retrograde counterparts.
The surface density profiles of the discs in these systems are shown in Fig. \ref{FIG:SIGMA-VS-R-GENERAL}.
As before the profiles are plotted in steps of $50\kilo\yr$.
We can see that all systems have most of their disc mass contained within $R_{\mathrm{max}} = 200\AU$ throughout their evolution.
With this value of $R_{\mathrm{max}}$, the values of $R_{\mathrm{av}}$ can be calculated from equation \eqref{EQ:WEIGHT-AVERAGE-R}.
In Fig. \ref{FIG:SIGMA-VS-R-GENERAL} we mark the values of $R_{\mathrm{av}}$ by blue squares on each evolving surface density profile.
We see that $R_{\mathrm{av}}$ marks the points at which the surface density begins to decline rapidly: i.e. it is the point within which the surface density is roughly constant.
We will return to the meaning of the red points later.

\begin{figure*}
    \centering
    \includegraphics[width=0.33\textwidth,angle=270]{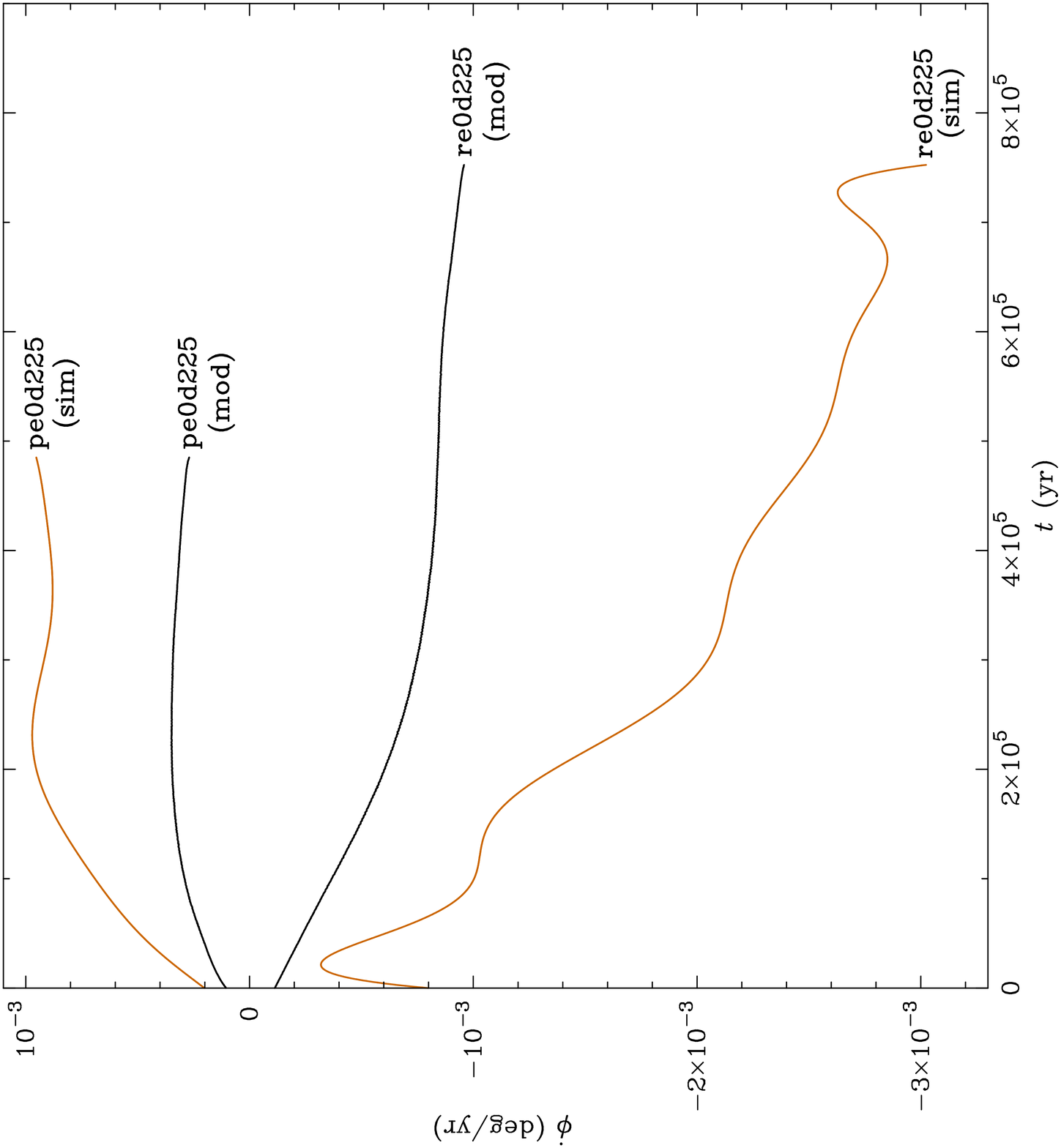}
    \includegraphics[width=0.33\textwidth,angle=270]{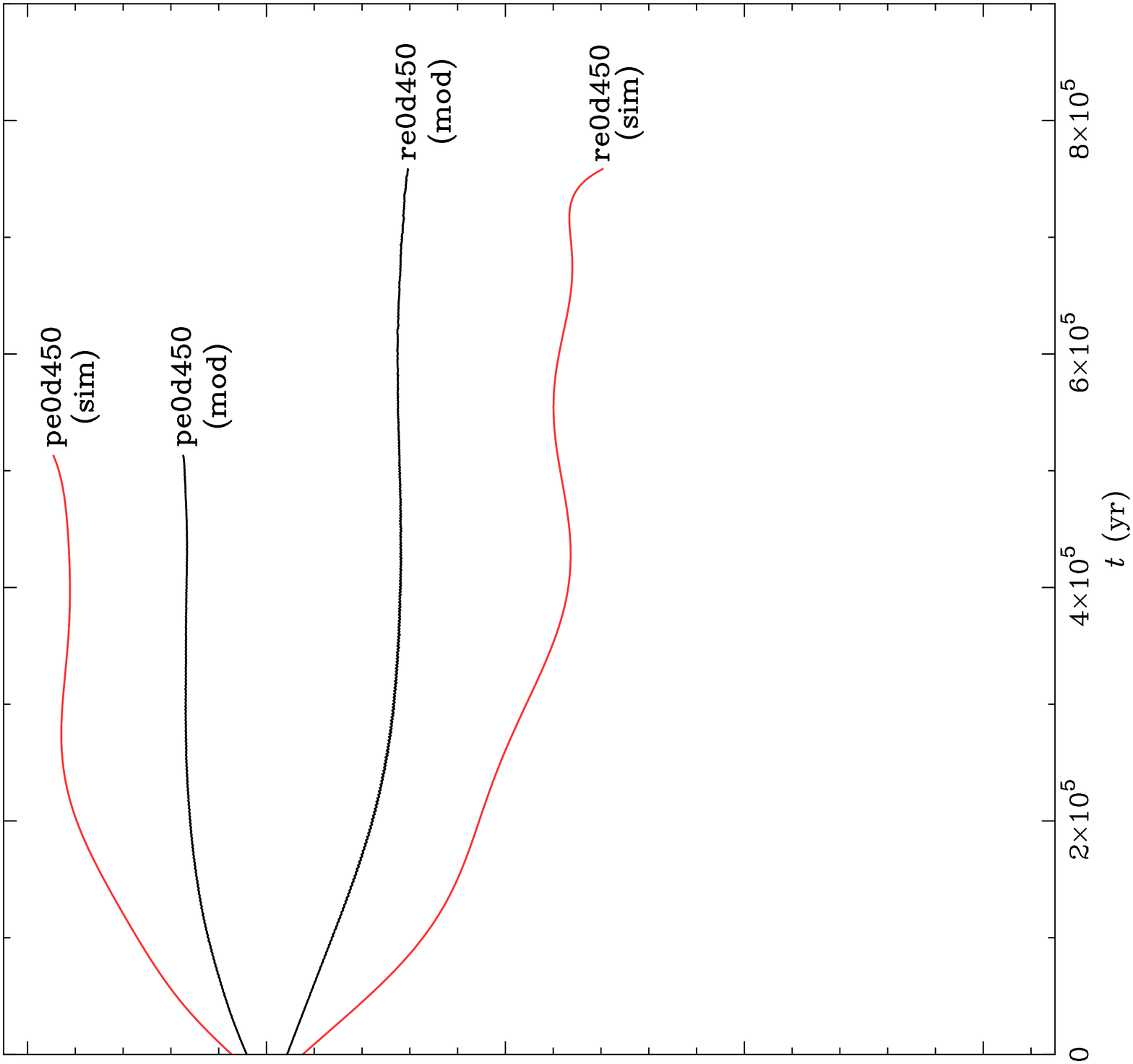}
    \includegraphics[width=0.33\textwidth,angle=270]{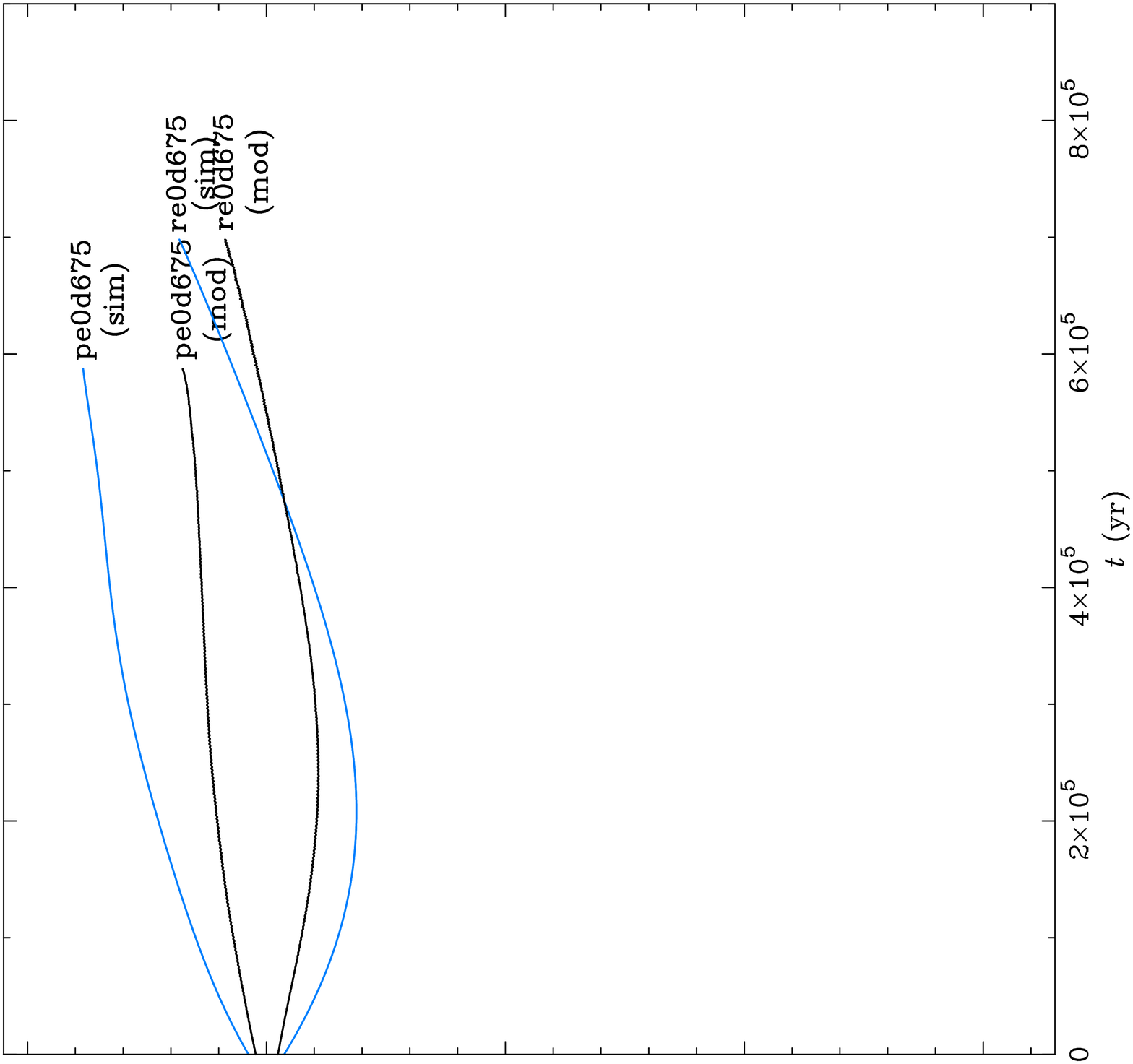}
    \caption{
        Precession rates ($\dot{\phi}$) with time ($t$) obtained from the model (mod) and the simulation results (sim) for initial misalignment angles of $22.5\degr$ (left panel), $45\degr$ (middle panel), and $67.5\degr$ (right panel).
    }
    \label{FIG:COMPARE-PHIDOT-VS-T}
\end{figure*}

Once we have the index $p$ and the radius $R_{\disc}$ for each simulation, the average precession rates can be obtained from equation \eqref{EQ:TIME-AVERAGED-PHI-2}.
Comparisons between the precession rates from the rigid-disc calculations and the simulations are shown in Fig. \ref{FIG:COMPARE-PHIDOT-VS-T}.
From left to right the panels show the prograde and retrograde simulations and rigid-disc calculations for $22.5\degr$, $45\degr$ and $67.5\degr$ misalignments.
In each panel, the solid black lines labelled with `mod' are the rigid-disc approximation and the lines with `sim' are the simulation results.

The rigid-disc model predicts the general behaviour seen in the simulations for all initial misalignment angles (comparing the models `mod', with the simulations `sim').
At moderate and high initial misalignments (middle and far right panels in Fig. \ref{FIG:COMPARE-PHIDOT-VS-T}), the models and simulations agree well but the magnitudes of the precession rate are under-estimated by the model by a factor of $\sim 2$.
At low initial misalignments (far left panel in Fig. \ref{FIG:COMPARE-PHIDOT-VS-T}), the model under-predicts the magnitudes of the precession rate by a factor of $2-3$.
Similar results are also found in $150\kilo$- and $600\kilo$-particles simulations of the same systems.

These results suggest that the rigid-disc model can describe the trend of the precession rate well, but underestimates the magnitude (especially when the initial misalignment is low).
The underestimation in magnitude is mainly due to our choice of $R_{\disc}$ (i.e. $R_{\mathrm{av}}$ from equation \eqref{EQ:WEIGHT-AVERAGE-R}) being somewhat arbitrary, and rather too low.

To obtain the `effective' radius ($R_{\mathrm{eff}}$) at which we would find the right magnitude for the precession rates, we simply use equation \eqref{EQ:TIME-AVERAGED-PHI-2} again with $\dot{\abrackets{\phi}}$ and other parameters taken from the result, but now solving the equation for $R_{\disc}$ instead.
For the selected systems, the values of $R_{\mathrm{eff}}$ which would match the simulation results are marked on the density profiles in Fig. \ref{FIG:SIGMA-VS-R-GENERAL} with red square symbols.

This is a rather back-to-front approach of fitting the models to the simulations and the simulations to the models.
However, Fig. \ref{FIG:SIGMA-VS-R-GENERAL} shows that {\em it is uncertainties in our disc radius estimates that cause the difference between the model predictions and the simulations, not any significant failure of the model.}
In most cases, the effective radius is larger than the average radius by a factor of $\sim 1.5-2$.
Similar results are also found in systems with higher eccentricities and different resolutions.

\subsection{Alignment rates}\label{SECT:ANALYSIS-ALIGNMENT-RATES}

Similarly to the precession rate above, we can compare the alignment rate from the simulations and the rigid-disc model in equation \eqref{EQ:TIME-AVERAGED-DELTA-1}.

For simplicity, let us consider a misaligned system with eccentricity $e = 0$.
The average alignment rate $\dot{\abrackets{\delta}}$ from equation \eqref{EQ:TIME-AVERAGED-DELTA-1} then reduces to
\begin{equation}\label{EQ:TIME-AVERAGED-DELTA-10}
    \dot{\abrackets{\delta}} \simeq -\frac{\dot{\abrackets{\phi}}\tan\delta}{2\pi}\int_{0}^{2\pi}\sin\sbrackets{2\rbrackets{\theta+\phi}}\dee{\theta}.
\end{equation}
To simplify further, we assume that $\phi$ changes linearly with time over an orbital period, i.e. $\phi \simeq \phi_{\circ}+\dot{\phi}\Delta{t}$.

For a circular orbit, where $\theta$ also changes linearly with time, $\Delta{t}$ can be written as $\Delta{t} = \theta/\dot{\theta}$, where $\theta$ ranges from $0$ to $2\pi$ and  $\dot{\theta}$ is constant.

The angle $\phi$ thus becomes $\phi \simeq \phi_{\circ}+(\dot{\phi}/\dot{\theta})\theta$.
Substituting $\phi$ in equation \eqref{EQ:TIME-AVERAGED-DELTA-10} gives us
\begin{equation}\label{EQ:TIME-AVERAGED-DELTA-11}
    \dot{\abrackets{\delta}} \simeq -\frac{\dot{\abrackets{\phi}}\tan\delta}{2\pi}\int_{0}^{2\pi}\sin\sbrackets{2\rbrackets{1+\frac{\dot{\phi}}{\dot{\theta}}}\theta+2\phi_{\circ}}\dee{\theta}.
\end{equation}
Integrating this equation gives
\begin{equation}\label{EQ:TIME-AVERAGED-DELTA-12}
    \dot{\abrackets{\delta}} \simeq -\frac{\dot{\abrackets{\phi}}\tan\delta}{4\pi\rbrackets{1+\dfrac{\dot{\phi}}{\dot{\theta}}}}\sbrackets{\cos(2\phi_{\circ})-\cos\rbrackets{2\phi_{\circ}+\frac{4\pi\dot{\phi}}{\dot{\theta}}}}.
\end{equation}
The rates $\dot{\phi}$ and $\dot{\theta}$ are obtained from the polynomial regressions of $\phi$ and $\theta$ respectively.

\begin{figure}
    \centering
    \includegraphics[angle=270,width=0.99\columnwidth]{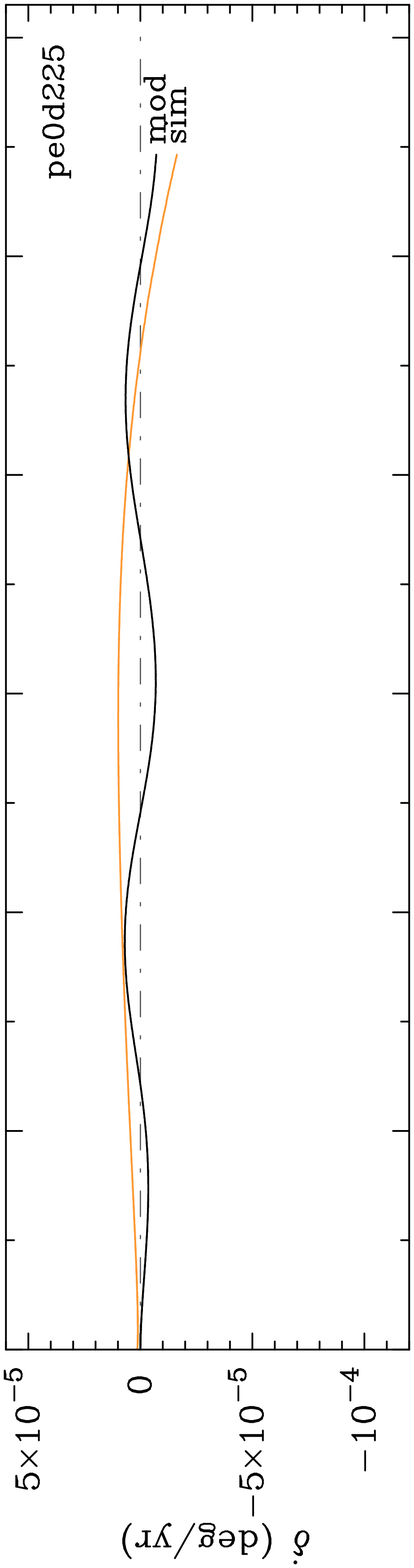}\\
    \includegraphics[angle=270,width=0.99\columnwidth]{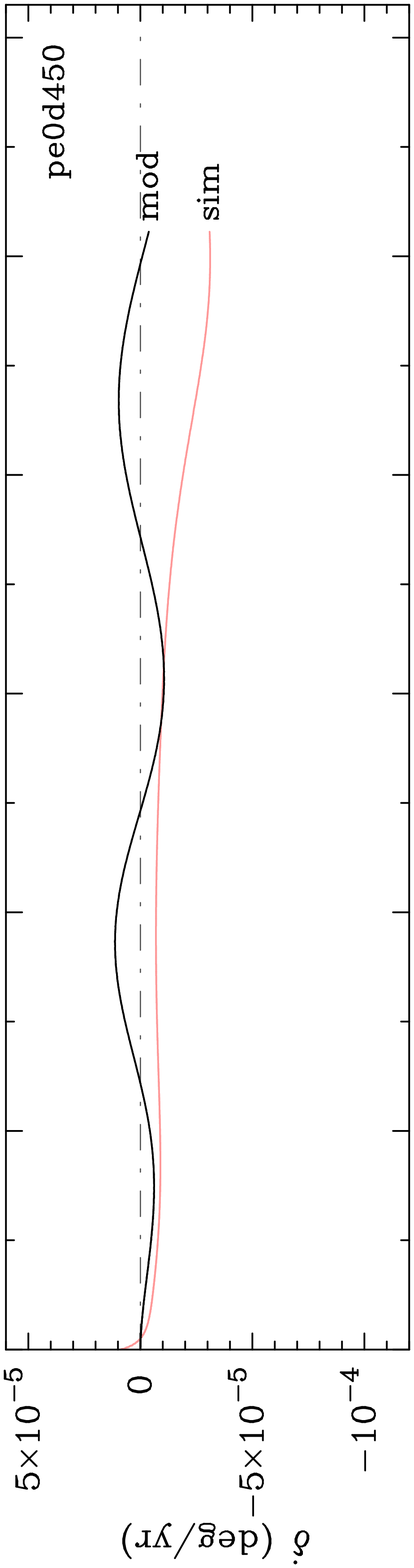}\\
    \includegraphics[angle=270,width=0.99\columnwidth]{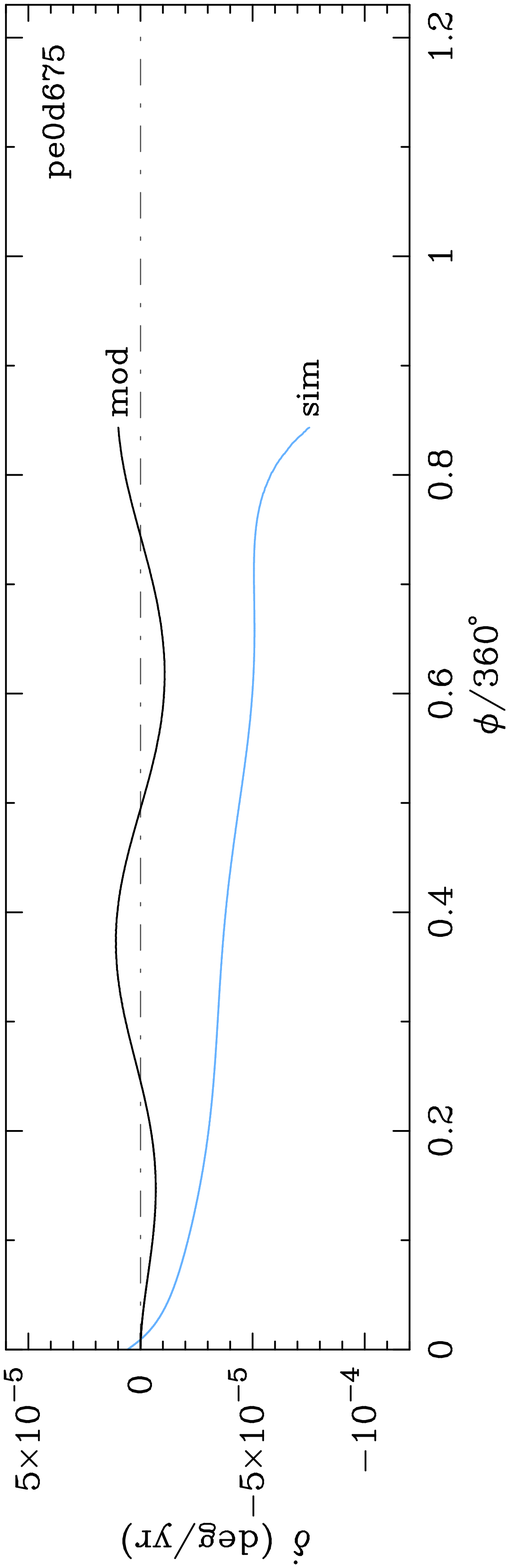}
    \caption{
        Alignment rates ($\dot{\delta}$) with time ($t$) obtained from the model (mod) and the simulation results (sim) for initial misalignment angles of $22.5\degr$ (top panel), $45\degr$ (middle panel), and $67.5\degr$ (bottom panel).
    }
    \label{FIG:COMPARE-DELDOT-VS-PHI-PROGRADE}
\end{figure}

The comparisons between $\dot{\abrackets{\delta}}$ from equation \eqref{EQ:TIME-AVERAGED-DELTA-12} and from the simulation result are shown in Fig. \ref{FIG:COMPARE-DELDOT-VS-PHI-PROGRADE}.
This figure shows $\dot{\abrackets{\delta}}$ against $\phi$, and here we only present prograde systems with initial misalignments of $22.5\degr$ (top), $45\degr$ (middle), and $67.5\degr$ (bottom).

In the rigid-disc predictions (black lines in Fig. \ref{FIG:COMPARE-DELDOT-VS-PHI-PROGRADE}), the alignment rate $\dot{\abrackets{\delta}}$ oscillates with a period of $\phi = 180\degr$, as suggested by equation \eqref{EQ:TIME-AVERAGED-DELTA-12}.
Given the low alignment rates in both the models and the simulations for initial misalignments of $22.5\degr$ and $45\degr$ (top and middle panels in Fig. \ref{FIG:COMPARE-DELDOT-VS-PHI-PROGRADE}), it is difficult to argue that the the model succeeds or fails to fit the simulated behaviour.
However, in the bottom panel, the model clearly fails to match the behaviour of the simulation.
The rigid-disc model predicts that the alignment rate should oscillate around zero, whilst the simulation clearly shows that it is growing with time.

In this case we find that the rigid-disc model fails, and that to calculate the alignment rate we require a model that accounts for the fluidity of the disc.

\subsection{Disc-companion misalignment in system with a non-rigid disc}\label{SECT:ANALYSIS-NON-RIGID-DISC-MODEL}

As we mentioned earlier in Section \ref{SECT:CHANGES-IN-BINARY-ORBITS}, the change in alignment between the disc and the companion orbit may be viewed as a consequence of the tidal torque and the encounter torque acting against each other.
We explain the mechanism in this subsection.

In a realistic misaligned system where the angular momentum $\bm{J}_{\disc}$ of the disc is not negligible compared to the angular momentum $\bm{J}_{\thebinary}$ of the binary, we have $\bm{J}_{\thebinary}$ and $\bm{J}_{\disc}$ precessing around the total angular momentum $\bm{J}_{\system}$ of the system, instead of $\bm{J}_{\disc}$ around $\bm{J}_{\thebinary}$ as in the rigid-disc model.
The three vectors also lie on the same plane.

\begin{figure}
    \centering
    \includegraphics[angle=0,width=0.99\columnwidth]{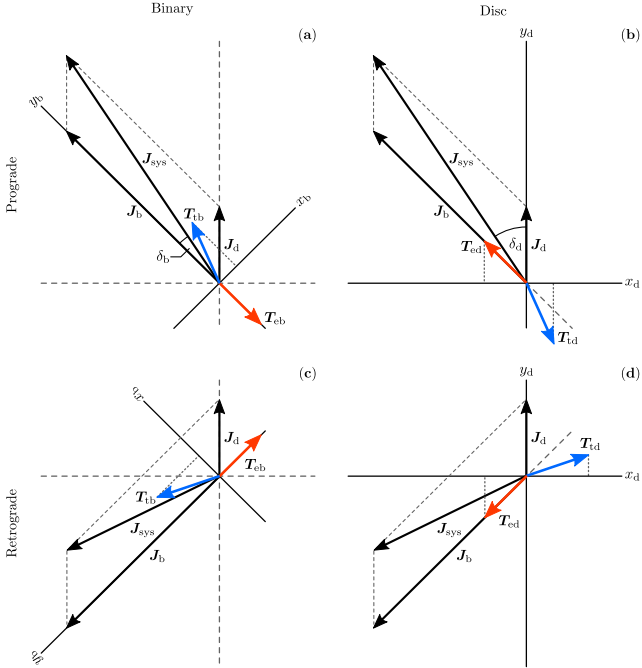}
    \caption{
        Orientations of the tidal torque $\bm{T}_{\tidal}$ (blue arrows) and the encounter torque $\bm{T}_{\encounter}$ (red arrows).
        The torques exerted on the binary orbit (panels (a) and (c)) have their own counterparts of the same magnitude but opposite in direction exerted on the disc (panels (b) and (d)).
        The misalignment angle is $\delta = \delta_{\thebinary}+\delta_{\disc}$, where $\delta_{\thebinary}$ is the angle between $\bm{J}_{\thebinary}$ and $\bm{J}_{\system}$, and $\delta_{\disc}$ is between $\bm{J}_{\disc}$ and $\bm{J}_{\system}$.
        Change in $\delta_{\thebinary}$ is due to the sum of torques projected on the $x_{\thebinary}$-axis, while change in $\delta_{\disc}$ is due to that projected on the $x_{\disc}$-axis.
    }
    \label{FIG:TORQUE-DIAGRAMS}
\end{figure}

For convenience, let us consider two new coordinate systems, $(x_{\thebinary},y_{\thebinary})$ for the binary and $(x_{\disc},y_{\disc})$ for the disc.
The $x_{\thebinary}-y_{\thebinary}$ plane and $x_{\disc}-y_{\disc}$ plane coincide.
The direction of the positive $y_{\thebinary}$-axis is defined by the vector $\bm{J}_{\thebinary}$ and the direction of the positive $y_{\disc}$-axis by the vector $\bm{J}_{\disc}$, as shown in Fig. \ref{FIG:TORQUE-DIAGRAMS}.

We know from Section \ref{SECT:CHANGES-IN-BINARY-ORBITS} that, on the binary orbit, the tidal torque $\bm{T}_{\tidal\thebinary}$ lies somewhere between $\bm{J}_{\thebinary}$ and $\bm{J}_{\disc}$ while the encounter torque $\bm{T}_{\encounter\thebinary}$ lies approximately opposite to $\bm{J}_{\thebinary}$.
In a prograde system, this can be illustrated in Fig. \ref{FIG:TORQUE-DIAGRAMS}(a) where $\bm{T}_{\tidal\thebinary}$ is represented (not to scale) by the blue arrow and $\bm{T}_{\encounter\thebinary}$ by the red arrow.
Similarly, the two torques in a retrograde system can be illustrated in Fig. \ref{FIG:TORQUE-DIAGRAMS}(c).
It should be noted that all the torques and angular momenta (except
$\bm{J}_{\system}$) we discuss here are time-averaged over an orbital
period (the instantaneous values oscillate somewhat).

Since the angular momentum $\bm{J}_{\system}$ of the system must be conserved, the counterpart torques of $\bm{T}_{\tidal\thebinary}$ and $\bm{T}_{\encounter\thebinary}$ exerted on the disc must be equal in magnitudes but opposite in directions.
That is, we have the tidal torque $\bm{T}_{\tidal\disc} = -\bm{T}_{\tidal\thebinary}$ and the encounter torque $\bm{T}_{\encounter\disc} = -\bm{T}_{\encounter\thebinary}$ exerted on the disc.
The two torques are illustrated (with the same colour code) in Fig. \ref{FIG:TORQUE-DIAGRAMS}(b) and (d) for prograde and retrograde systems respectively.

Now we have the net torque $\bm{T}_{\thebinary} = \bm{T}_{\tidal\thebinary}+\bm{T}_{\encounter\thebinary}$ as the net rate of change of $\bm{J}_{\thebinary}$ and $\bm{T}_{\disc} = \bm{T}_{\tidal\disc}+\bm{T}_{\encounter\disc}$ as the net rates of change of $\bm{J}_{\disc}$.
However, it is the components of $\bm{T}_{\thebinary}$ on the $x_{\thebinary}$-axis and $\bm{T}_{\disc}$ on the $x_{\disc}$-axis that actually cause a change in the misalignment angle $\delta$.

Let us consider the misalignment angle $\delta$ as the sum of the angles $\delta_{\thebinary}$, between $\bm{J}_{\thebinary}$ and $\bm{J}_{\system}$ (see Fig. \ref{FIG:TORQUE-DIAGRAMS}(a)), and $\delta_{\disc}$, between $\bm{J}_{\disc}$ and $\bm{J}_{\system}$ (see Fig. \ref{FIG:TORQUE-DIAGRAMS}(b)).
That is, $\delta = \delta_{\thebinary}+\delta_{\disc}$.
The change in $\delta_{\thebinary}$ is due to the component of $\bm{T}_{\thebinary}$ on the $x_{\thebinary}$-axis, as well as the change in $\delta_{\disc}$ is due to the component of $\bm{T}_{\disc}$ on the $x_{\disc}$-axis.

We can see from Fig. \ref{FIG:TORQUE-DIAGRAMS}(a) and (c) that the component of $\bm{T}_{\thebinary}$ on the $x_{\thebinary}$-axis is mostly from $\bm{T}_{\tidal\thebinary}$, since $\bm{T}_{\encounter\thebinary}$ is approximately perpendicular to the $x_{\thebinary}$-axis.
The change in $\delta_{\thebinary}$ therefore depends only on the tidal torque $\bm{T}_{\tidal\thebinary}$.
In contrast, the component of $\bm{T}_{\disc}$ on the $x_{\disc}$-axis (and thus the change in $\delta_{\disc}$) depends on both $\bm{T}_{\tidal\disc}$ and $\bm{T}_{\encounter\disc}$, see Fig. \ref{FIG:TORQUE-DIAGRAMS}(b) and (d).

\begin{figure}
    \centering
    \includegraphics[angle=270,width=0.99\columnwidth]{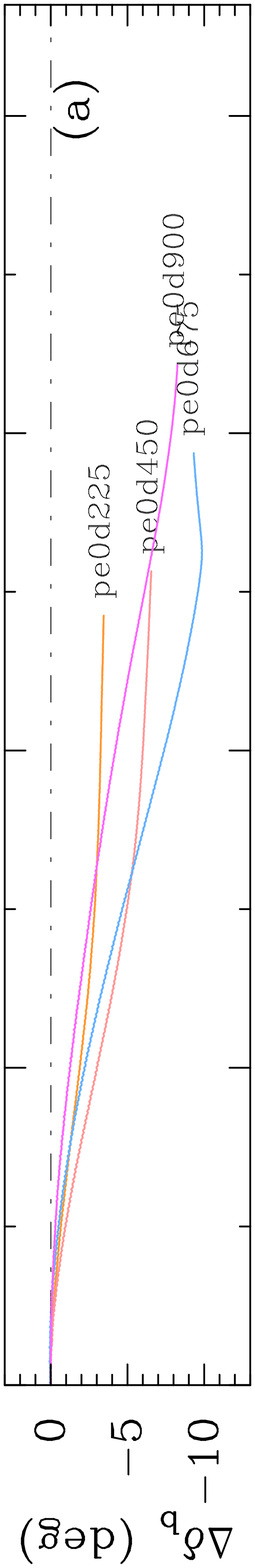}\\
    \includegraphics[angle=270,width=0.99\columnwidth]{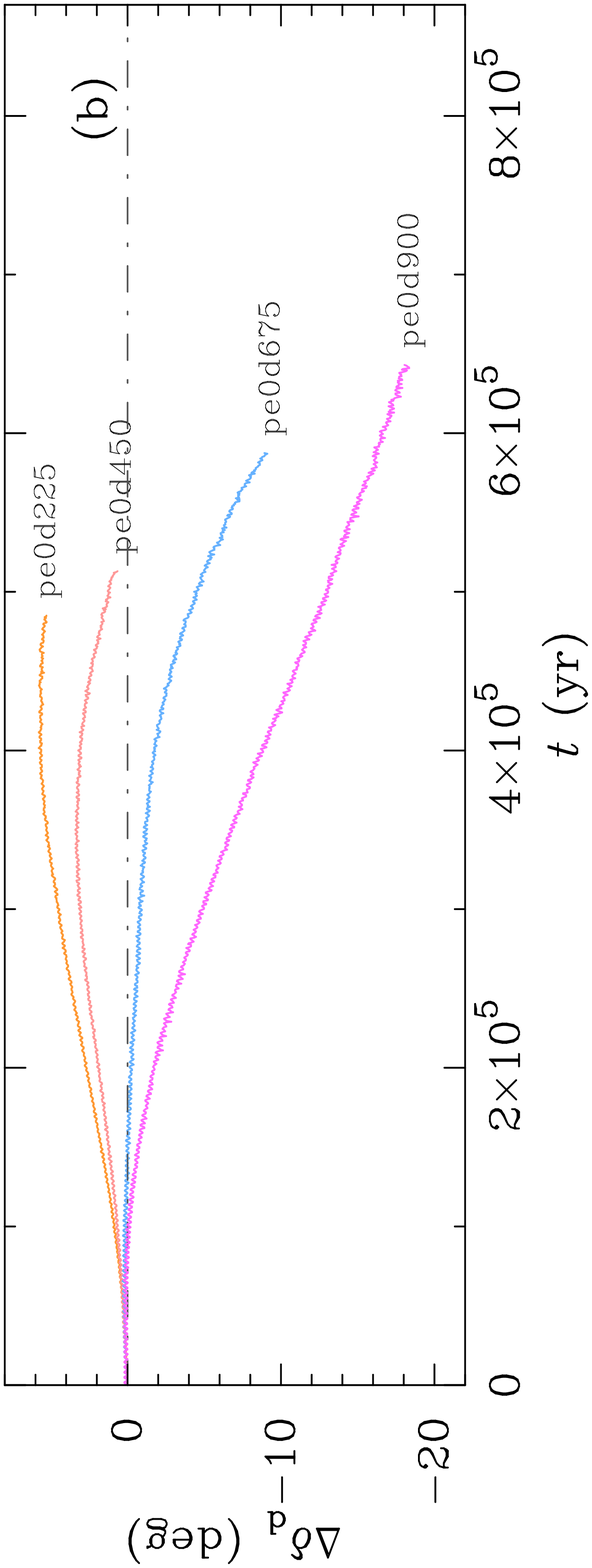}
    \caption{
        Change of $\delta_{\thebinary}$ (top panel) and $\delta_{\disc}$ (bottom panel) with time ($t$) in selected prograde systems.
    }
    \label{FIG:COMPARE-DDELTA-VS-T-PROGRADE}
\end{figure}

\begin{figure}
    \centering
    \includegraphics[angle=270,width=0.99\columnwidth]{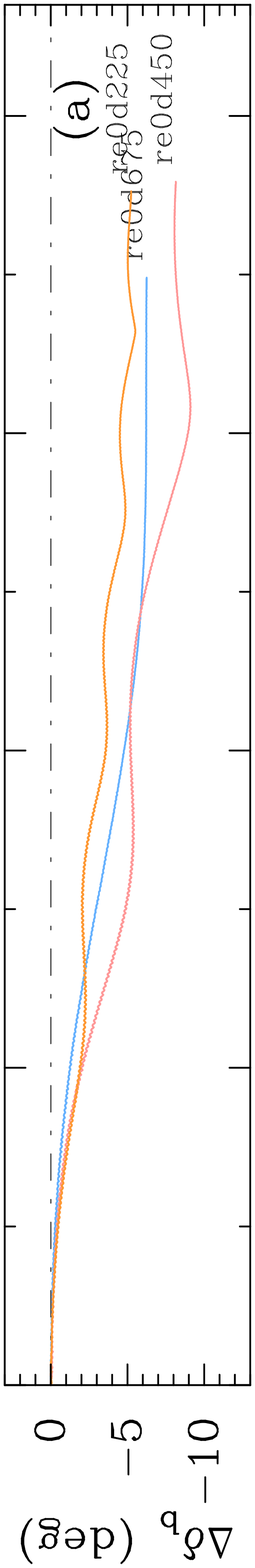}\\
    \includegraphics[angle=270,width=0.99\columnwidth]{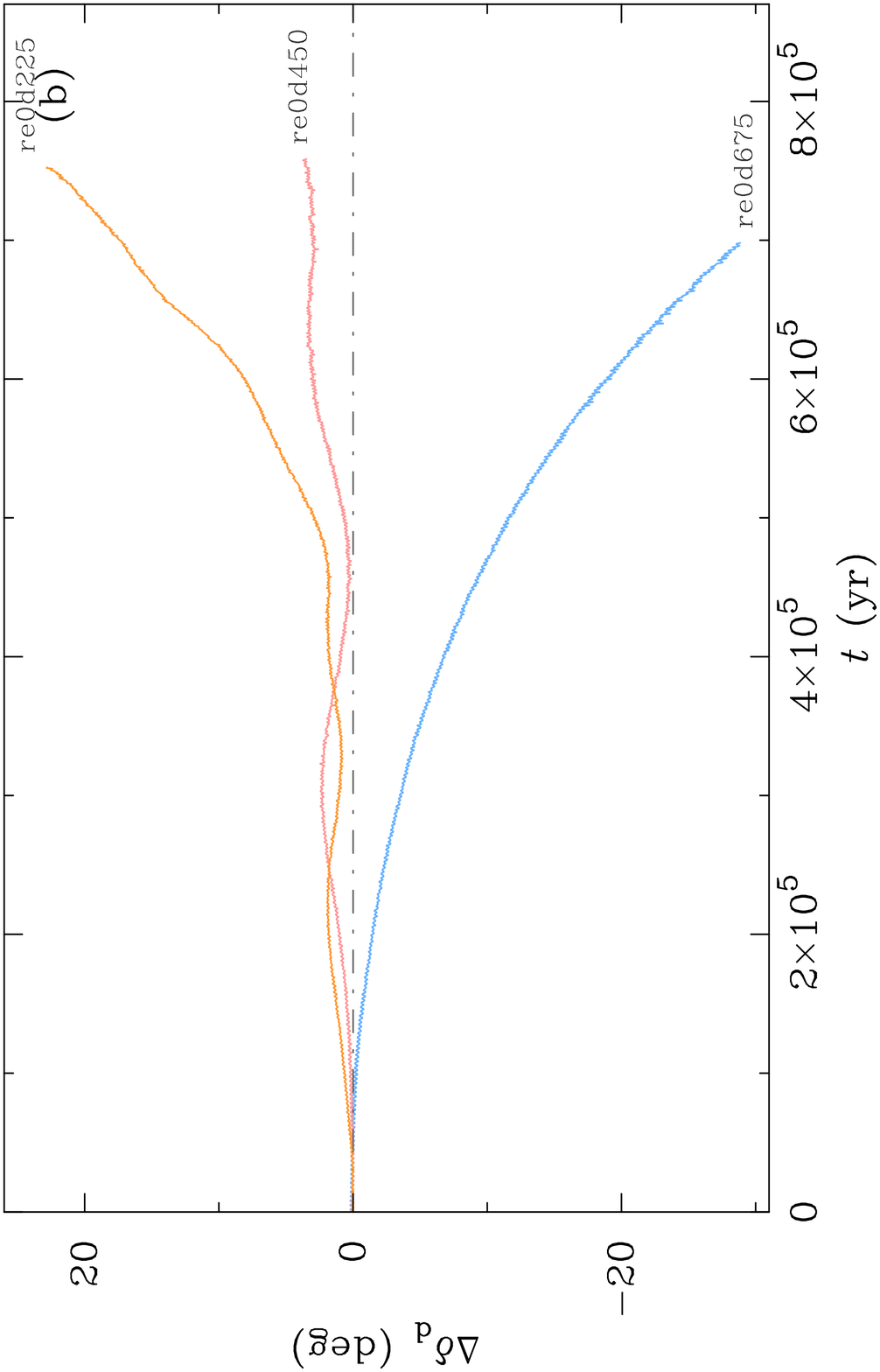}
    \caption{
        Change of $\delta_{\thebinary}$ (top panel) and $\delta_{\disc}$ (bottom panel) with time ($t$) in selected retrograde systems.
    }
    \label{FIG:COMPARE-DDELTA-VS-T-RETROGRADE}
\end{figure}

To explain how $\delta_{\thebinary}$ and $\delta_{\disc}$ are changed by the associated torques, we examine the differences between the angles and their initial values ($\Delta\delta_{\thebinary}$ and $\Delta\delta_{\disc}$) of some prograde systems in Fig. \ref{FIG:COMPARE-DDELTA-VS-T-PROGRADE}, and retrograde systems in Fig. \ref{FIG:COMPARE-DDELTA-VS-T-RETROGRADE}.
Note that the net difference in misalignment angle is $\Delta\delta = \delta-\delta_{\circ} = \Delta\delta_{\thebinary}+\Delta\delta_{\disc}$.
Using what we also know from Section \ref{SECT:CHANGES-IN-BINARY-ORBITS} that the tidal torque dominates the encounter torque when $\delta$ is close to $0\degr$ or $180\degr$, and vice versa when $\delta$ is close to $90\degr$, we can explain the changes in Fig. \ref{FIG:COMPARE-DDELTA-VS-T-PROGRADE} and \ref{FIG:COMPARE-DDELTA-VS-T-RETROGRADE} as follows.

{\bf The change in $\delta_{\thebinary}$.}
In the binary frame of reference in Fig. \ref{FIG:TORQUE-DIAGRAMS}(a) and (c) where the encounter torque $\bm{T}_{\encounter\thebinary}$ is perpendicular to the $x_{\thebinary}$-axis, the change in $\delta_{\thebinary}$ is only due to the component of the tidal torque $\bm{T}_{\tidal\thebinary}$ on the positive $x_{\thebinary}$-axis.
The torque will bring $\bm{J}_{\thebinary}$ towards $\bm{J}_{\system}$, decreasing $\delta_{\thebinary}$ from its initial value.
This results in the negative value of $\Delta\delta_{\thebinary}$ in both prograde and retrograde systems, as shown in Fig. \ref{FIG:COMPARE-DDELTA-VS-T-PROGRADE}(a) and \ref{FIG:COMPARE-DDELTA-VS-T-RETROGRADE}(a).
Note that the component of $\bm{T}_{\tidal\thebinary}$ on the $x_{\thebinary}$-axis depends on both the magnitude of $\bm{T}_{\tidal\thebinary}$ itself, which increases with decreasing $\delta$, and the cosine value of the angle between $\bm{T}_{\tidal\thebinary}$ and the $x_{\thebinary}$-axis, which increases with increasing $\delta$.
This makes the curves in Fig. \ref{FIG:COMPARE-DDELTA-VS-T-PROGRADE}(a) become more negative from \texttt{pe0d225} to \texttt{pe0d450} and then less negative from \texttt{pe0d675} to \texttt{pe0d900}; the turning point is at some angle between $45\degr$ and $67.5\degr$.
Similar trends are also found in retrograde systems as shown in Fig. \ref{FIG:COMPARE-DDELTA-VS-T-RETROGRADE}(a).

{\bf The change in $\delta_{\disc}$.}
In the disc's frame of reference in Fig. \ref{FIG:TORQUE-DIAGRAMS}(b) and (d), the net torque on the $x_{\disc}$-axis depends on both tidal torque and encounter torque.
In systems with $\delta$ close to $0\degr$ or $180\degr$, where $\bm{T}_{\tidal\disc}$ is almost on the $y_{\thebinary}$-axis and having magnitude greater than $\bm{T}_{\encounter\disc}$, the component of $\bm{T}_{\tidal\disc}$ dominates that of $\bm{T}_{\encounter\disc}$ on the $x_{\disc}$-axis.
The net torque then points along the positive $x_{\disc}$-axis, turning $\bm{J}_{\disc}$ away from $\bm{J}_{\system}$.
The value of $\delta_{\disc}$ is therefore increased from its initial value (positive $\Delta\delta_{\disc}$), as we can see for \texttt{pe0d225} in Fig. \ref{FIG:COMPARE-DDELTA-VS-T-PROGRADE}(b) and \texttt{re0d225} in \ref{FIG:COMPARE-DDELTA-VS-T-RETROGRADE}(b).

In more misaligned systems, on the other hand, the magnitude of $\bm{T}_{\encounter\disc}$ increases while the magnitude of $\bm{T}_{\tidal\disc}$ decreases.
The direction of $\bm{T}_{\tidal\disc}$ also turns away from the $y_{\thebinary}$-axis, making its component on the $x_{\disc}$-axis even smaller.
In highly misaligned systems, the component of $\bm{T}_{\encounter\disc}$ on the negative $x_{\disc}$-axis dominates the component of $\bm{T}_{\tidal\disc}$ on the positive $x_{\disc}$-axis.
The net torque on the negative $x_{\disc}$-axis will then turn $\bm{J}_{\disc}$ towards $\bm{J}_{\system}$, decreasing the value of $\delta_{\disc}$.
This situation occurs in \texttt{pe0d675} and \texttt{pe0d900} in Fig. \ref{FIG:COMPARE-DDELTA-VS-T-PROGRADE}(b), and \texttt{re0d675} in Fig. \ref{FIG:COMPARE-DDELTA-VS-T-RETROGRADE}(b).

\section{Conclusions}\label{SECT:SUMMARY}

We have performed a number of simulations of circumprimary discs in misaligned binary systems.
The companion starts misaligned to the disc at $22.5\degr$, $45\degr$ and $67.5\degr$ in both prograde and retrograde orbits with eccentricities from $0$ to $0.6$.
We assume that the disc and central star begin with aligned rotation.
We then compare the outcome of our simulations to the analytic rigid-disc approximation.

Our goal is to examine how the relative misalignments of the companion and disc, and disc and primary star change with time.

We find that a misaligned companion will misalign the circumprimary disc with respect to the primary star.
The degree and speed of this misalignment depends on the initial misalignment of the disc and companion as well as the orbital direction of the companion.

Firstly, we have shown that the rigid-disc model can describe the precession rate of the disc.
However, we find that the rigid-disc model fails to describe the change in alignment between the disc and the companion star (Section \ref{SECT:ANALYSIS-ALIGNMENT-RATES}).
The failure of the rigid-disc model implies that the alignment process is associated with torques exerted on the disc as fluid body.
The most plausible torques are the tidal torque (which dominates in systems with initial misalignments near $0\degr$ or $180\degr$), and the encounter torque (dominating in systems with initial misalignments near $90\degr$).
The tidal torque tends to make the disc-companion system more misaligned, while the encounter torque tends to align them.
Although we have not provided any mathematical description for these torques, our schematic description presented in Section \ref{SECT:ANALYSIS-NON-RIGID-DISC-MODEL} seems to be consistent with the results.

Secondly, our simulation results suggest that complete alignment between the disc and the binary orbit in misaligned systems may never be achieved within the disc lifetime.
This is because (1) the change in alignment is a slow process, especially in systems with $\delta \ll 90\degr$, and (2) there is an anti-alignment process (due to the tidal torque), in system with low $\delta$, preventing the systems from being aligned.
Therefore, misaligned systems will eventually stay misaligned.

Finally, we have also shown that precession is an efficient process to misalign the rotational axis of the disc from the spin axis of the host star.
Precession can make the star-disc misalignment angle ($\psi$) reach a value near $2\delta_{\circ}$, where $\delta_{\circ}$ is the initial disc-binary misalignment angle.
The disc can then be highly misaligned or even temporarily retrograde in systems with high $\delta_{\circ}$, e.g. $\delta_{\circ} > 45\degr$ (Section \ref{SECT:STAR-DISC-MISALIGNMENT-ANGLE}).

The change in the misalignment angles between the primary, disc and companion may be important in explaining the origin of (some) spin-orbit misaligned exoplanetary systems \citep{Batygin:2012}.

\section*{Acknowledgements}

We thank the anonymous referee for useful comments that helped improve the manuscript.
KR and SPG acknowledge support from a Leverhulme Trust international network grant (IN-2013-017).
KR also acknowledges support from Faculty of Science Research Fund (2013), Prince of Songkla University.
Simulations in this work have been performed on Iceberg, the High
Performance Computing server at the University of Sheffield.




\bibliographystyle{mnras}
\bibliography{references} 




\appendix

\section{Particle distribution in an SPH disc}

An SPH disc can be constructed by distributing gas particles in a volume constrained by equation \eqref{EQ:SURFACE-DENSITY-PROFILE-1} and \eqref{EQ:TEMPERATURE-PROFILE-1} \citep[e.g.][]{Stamatellos:Whitworth:2008}.
It is convenient to calculate the distribution in cylindrical coordinates ($R,\varphi,z$) and then transform to Cartesian coordinates ($x,y,z$) which are used in the simulations.

The radial distribution of particles can be constructed by randomly sampling the mass distribution $f_{R} = M(R)/M_{\disc}$, where $M(R)$ is the mass of the disc within radius $R$ from the central star.
The mass $M(R)$ can be obtained from integrating $\dee{M} = 2\pi\Sigma R\dee{R}$ from $R = R_{\mathrm{in}}$ to $R$.
One can rearrange the terms and find that
\begin{equation}
    R = \sbrackets{\frac{3}{4\pi}\frac{M_{\disc}}{\Sigma_{1}}f_{R}+R_{\mathrm{in}}^{3/2}}^{2/3}.
\end{equation}
Drawing a set of random numbers $f_{R}$ uniform in $0 \leq f_{R} \leq 1$, we then have the radial distribution of particles for the disc.
Similarly, the azimuthal distribution can be constructed by randomly sampling the angle $f_{\varphi} = \varphi/2\pi$.
This is the standard way of sampling to match a particular mass distribution.

The vertical distribution, on the other hand, is rather more complicated, we construct it from the fraction of surface densities
\begin{equation}\label{EQ:VERTICAL-DISTRIBUTION1}
    f_{z}   = \frac{\int_{-z}^{z}\dee{\Sigma}}{\int_{-\infty}^{\infty}\dee{\Sigma}} = \frac{\int_{-z}^{z}\rho\dee{z}}{\int_{-\infty}^{\infty}\rho\dee{z}},
\end{equation}
where $\rho = \rho(R,z)$ is the volume density.

To find an expression for $\rho(R,z)$, let us consider the balance of vertical accelerations at the vertical scale height $z_{\circ} = z_{\circ}(R)$ of a thin disc ($z_{\circ} \ll R$):
\begin{equation}\label{EQ:VERTICAL-ACCELERATIONS1}
    \frac{GM_{\primary}z_{\circ}}{R^{3}}+2\pi{}G\Sigma = -\frac{1}{\rho}\frac{\partial{P}}{\partial{z}}\bigg\vert_{{z = z_{\circ}}},
\end{equation}
where $G$ is the gravitational constant and $P = P(R,z)$ is the local pressure.
The first and second terms on the LHS are the gravitational accelerations due to the central star and the self gravity between disc particles (estimated by Gauss's law for gravity) respectively.
The term on the RHS is the vertical hydrostatic acceleration in the disc.

In an isothermal approximation, the pressure $P$ may be written in terms of density $\rho$ and sound speed $c_{\secondary}$ (or temperature $T$) as
\begin{equation}\label{EQ:ISOTHERMAL-PRESSURE}
    P(R,z) = \rho(R,z)c_{\secondary}^{2}(R) = \rho(R,z)\sbrackets{\frac{k_{\textsc{b}}{}T(R)}{\bar{\mu}m_{\scriptscriptstyle\mathrm{H}}}},
\end{equation}
where $k_{\textsc{b}}$ is the Boltzmann constant, $\bar{\mu} = 2.35$ the mean molecular weight, and $m_{\scriptscriptstyle\mathrm{H}}$ the hydrogen mass.

In a system with $M_{\disc} \ll M_{\primary}$, the second term on the LHS of equation \eqref{EQ:VERTICAL-ACCELERATIONS1} is very small compared to the first term.
In this case, the approximate expression for $\rho(R,z)$ can be obtained by neglecting the second term (for a moment) and considering $z_{\circ}$ as $z$.
We see that $\partial{\rho}/\partial{z} \propto -z\rho$.
This means that the local density is a Gaussian function of $z$, i.e.
\begin{equation}\label{EQ:GAUSSIAN-FUNCTION1}
    \rho(R,z) = \rho(R,0)\encounter^{-(bz/z_{\circ})^{2}},
\end{equation}
where $\rho(R,0)$ is the volume density at the disc midplane and $b$ is an arbitrary constant.
By using $\rho$ from equation \eqref{EQ:GAUSSIAN-FUNCTION1}, the vertical scale height ($z_{\circ}$) obtained from solving equation \eqref{EQ:VERTICAL-ACCELERATIONS1} can be written as
\begin{equation}\label{EQ:VERTICAL-SCALE-HEIGHT}
    z_{\circ} \simeq -\frac{\pi\Sigma(R)R^{3}}{M_{\primary}}+\sbrackets{\rbrackets{\frac{\pi\Sigma(R)R^{3}}{M_{\primary}}}^{2}+\frac{2b^{2}R^{3}c_{\secondary}^{2}(R)}{GM_{\primary}}}^{1/2}.
\end{equation}

In order to obtain the vertical distribution from equation \eqref{EQ:VERTICAL-DISTRIBUTION1}, however, we employ an integrable Gaussian-like function
\begin{equation}\label{EQ:GAUSSIAN-FUNCTION2}
    \rho(R,z) \simeq \rho(R,0)\mathrm{sech}^{2}(bz/z_{\circ})
\end{equation}
instead of the exact Gaussian function in equation (\ref{EQ:GAUSSIAN-FUNCTION1}).
By using this function for the density in equation \eqref{EQ:VERTICAL-DISTRIBUTION1}, we have
\begin{equation}
     z = \frac{z_{\circ}}{b}\mathrm{tanh}^{-1}\rbrackets{f_{z}},
\end{equation}
where $z_{\circ}$ is obtained from equation \eqref{EQ:VERTICAL-SCALE-HEIGHT}.
The value of $b$ defines the initial thickness of the disc: the bigger the value, the thicker the disc.
For our simulations, we simply use $b = 1$.
Finally, the random number $f_{z}$ in this case is $-1 < f_{z} < 1$ (exclusive).


\bsp    
\label{lastpage}
\end{document}